%% file: 2D_Pumping_arXiv.tex
\documentclass[%
	10pt,
	showpacs,%
	amsmath,%
	amssymb,%
	aps,%
	twocolumn,%
	prl,%
	nofootinbib%
]{revtex4-1}

\usepackage[english]{babel}
\usepackage[utf8]{inputenc}
\usepackage[T1]{fontenc}
\usepackage{csquotes}

\usepackage[colorlinks=true,citecolor=blue,linkcolor=black,urlcolor=blue]{hyperref}

\usepackage{graphicx}

\usepackage[cspex,bbgreekl]{mathbbol}
\newcommand{\ket}[1]{\left| #1 \right>} 
\newcommand{\braket}[2]{\left< #1 | #2 \right>} 

\newcommand{\sub}[1]{_{\mathrm{#1}}}
\newcommand{\supscr}[1]{^{\mathrm{#1}}}
\newcommand{\fig}[1]{Fig.~\ref{#1}}
\newcommand{\extfig}[1]{Fig.~\ref{#1}}
\newcommand{\eq}[1]{Eq.~(\ref{#1})}

\newcommand {\grsim} {\ {\raise-.5ex\hbox{$\buildrel>\over\sim$}}\ }
\newcommand {\lessim} {\ {\raise-.5ex\hbox{$\buildrel<\over\sim$}}\ }

\newcommand{\invisiblesection}[1]{%
  \phantomsection%
  \stepcounter{section}%
  \addcontentsline{toc}{section}{\protect\numberline{\thesection}#1}%
  }

\begin{document}

\title{Exploring 4D Quantum Hall Physics with a 2D Topological Charge Pump}

\author{Michael Lohse$^{1,2}$, Christian Schweizer$^{1,2}$, Hannah M. Price$^{3}$, Oded Zilberberg$^{4}$ \& Immanuel Bloch$^{1,2}$}

\affiliation{
\vspace{0.5em}
\mbox{$^{1}$\,Fakult\"at f\"ur Physik, Ludwig-Maximilians-Universit\"at,~Schellingstra\ss e 4, 80799 M\"unchen, Germany}\\
\mbox{$^{2}$\,Max-Planck-Institut f\"ur Quantenoptik, Hans-Kopfermann-Stra\ss e 1, 85748 Garching, Germany}\\
\mbox{$^{3}$\,INO-CNR BEC Center \& Dipartimento di Fisica, Universit\`a di Trento, Via Sommarive 14, 38123 Povo, Italy}\\
\mbox{$^{4}$\,Institut f\"ur Theoretische Physik, ETH Z\"{u}rich, Wolfgang-Pauli-Stra\ss e 27, 8093 Z\"urich, Switzerland}\\
}

\maketitle

\invisiblesection{Abstract}
{\bf The discovery of topological states of matter has profoundly augmented our understanding of phase transitions in physical systems. Instead of local order parameters, topological phases are described by global topological invariants and are therefore robust against perturbations. A prominent example thereof is the two-dimensional integer quantum Hall effect \cite{Klitzing:1980}. It is characterized by the first Chern number which manifests in the quantized Hall response induced by an external electric field \cite{Thouless:1982}. Generalizing the quantum Hall effect to four-dimensional systems leads to the appearance of a novel non-linear Hall response that is quantized as well, but described by a 4D topological invariant -- the second Chern number \cite{Yang:1978, Zhang:2001}. Here, we report on the first observation of a bulk response with intrinsic 4D topology and the measurement of the associated second Chern number. By implementing a 2D topological charge pump with ultracold bosonic atoms in an angled optical superlattice, we realize a dynamical version of the 4D integer quantum Hall effect \cite{Thouless:1983, Kraus:2013}. Using a small atom cloud as a local probe, we fully characterize the non-linear response of the system by in-situ imaging and site-resolved band mapping. Our findings pave the way to experimentally probe higher-dimensional quantum Hall systems, where new topological phases with exotic excitations are predicted \cite{Zhang:2001}.}

\begin{figure}[t!]
\vspace{0.75mm}
\includegraphics[width=\linewidth]{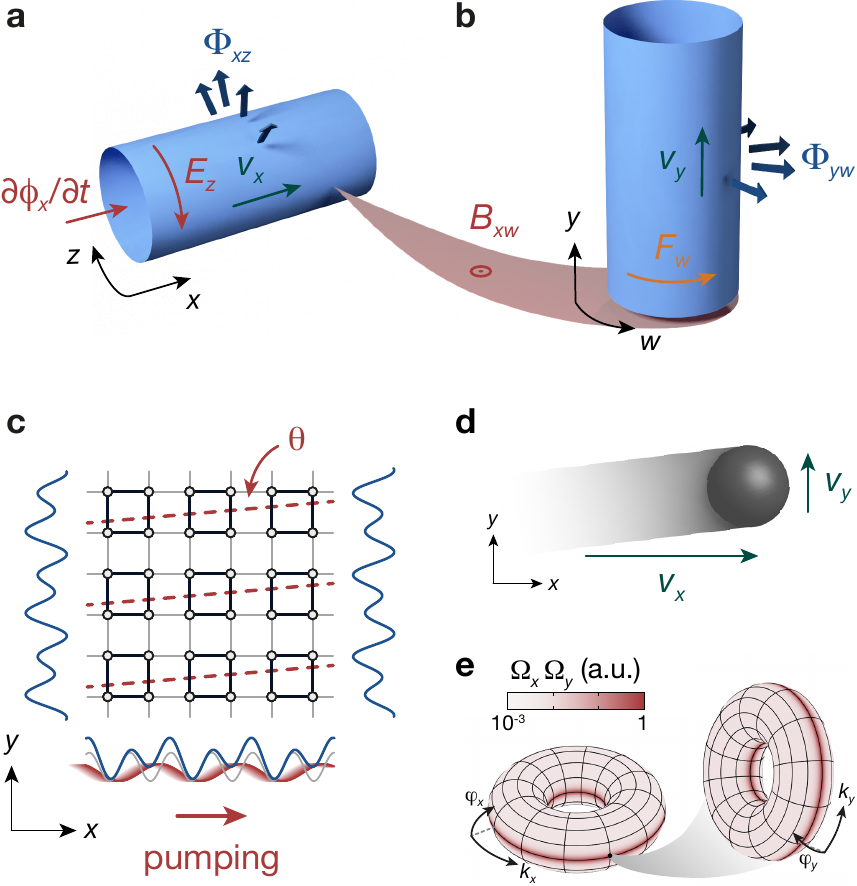}
\vspace{0.75mm} \caption{Four-dimensional quantum Hall (QH) system and corresponding 2D topological charge pump. 
\textbf{(a)} A 2D QH system on a cylinder pierced by a uniform magnetic flux $\Phi_{xz}$. Threading a magnetic flux $\phi_x$ through the cylinder creates an electric field $E_z$ on the surface, resulting in a linear Hall response along $x$ with velocity $v_x$. 
\textbf{(b)} A 4D QH system can be composed of two 2D QH systems in the $xz$- and $yw$-planes. A weak magnetic perturbation $B_{xw}$ in the $xw$-plane couples the two systems and generates a Lorentz force $F_w$ for particles moving along $x$. This induces an additional non-linear Hall response in the $y$-direction with velocity $v_y$. 
\textbf{(c)} A dynamical version of the 4D QH system can be realized with a topological charge pump in a 2D superlattice (blue potentials). Such a superlattice is created by superimposing two lattices with periodicities $d\sub{s}$ (grey) and $d\sub{l} > d\sub{s}$ (red) along both $x$ and $y$, depicted here for $d\sub{l} = 2 d\sub{s}$ as in the experiment. The black circles show the lattice sites formed by the potential minima and the black (grey) lines indicate strong (weak) tunnel coupling between neighbouring sites. The system is modulated periodically by adiabatically moving the long lattice along $x$, mimicking the perturbing electric field $E_z$ in the 4D model. The magnetic perturbation $B_{xw}$ maps onto a small tilt angle $\theta$ of the long lattice with respect to the short lattice along $y$. In this case, the shape of the double wells along $y$ depends on the position along $x$. The dashed red lines indicate the potential minima of the tilted long lattice. 
\textbf{(d)} The pumping gives rise to a motion of the atom cloud in the $x$-direction, corresponding to the quantized linear response of a 2D QH system. For non-zero $\theta$, the two orthogonal axes are coupled, leading to an additional quantized non-linear response with 4D topology in the perpendicular $y$-direction. 
\textbf{(e)} The velocity of the non-linear response is determined by the product of the Berry curvatures $\Omega^x \Omega^y$ (see Supplementary Information), depicted here for the lowest subband with $d\sub{l} = 2 d\sub{s}$ and lattice depths as in \fig{fig:3}. The left (right) torus shows a cut at $k_y = 0$, $\varphi_y = \pi/2$ ($k_x = \pi/(2d\sub{l})$, $\varphi_x = \pi/2$) through the generalized 4D Brillouin zone spanned by $k_x$, $\varphi_x$, $k_y$ and $\varphi_y$. 
\label{fig:1}}
\end{figure}

\invisiblesection{4D Quantum Hall Effect}
Topology, originally a branch of mathematics, has become an important concept in different fields of physics, ranging from particle physics \cite{Yang:1954, Yang:1978} to solid state physics \cite{Qi:2011} and quantum computation \cite{Nayak:2008}. In this context, a hallmark achievement was the discovery of the 2D integer quantum Hall (QH) effect \cite{Klitzing:1980}. It demonstrated that the Hall conductance, in response to an electric field $\mathbf{E}$, is quantized for charged particles moving in 2D in the presence of a perpendicular magnetic field. In a cylindrical geometry, following Laughlin's gedankenexperiment, $\mathbf{E}$ can be generated by varying the magnetic flux $\phi_x(t)$ along the axis of the cylinder \cite{Laughlin:1981} (\fig{fig:1}a). Due to the interplay of the perpendicular magnetic field and the induced electric field $E_z$, the particles start moving in the $x$-direction. This gives rise to the quantized Hall response through an integer number of particles that is transported from one edge to the other per flux quantum threaded through the cylinder. This quantized response is characterized by an integer topological invariant, the first Chern number \cite{Thouless:1982}.

Dimensionality plays a crucial role for topological phases and many intriguing states have recently been discovered in 3D systems, e.g.~Weyl semimetals \cite{Lu:2015, Xu:2015} and 3D topological insulators \cite{Hsieh:2008}. Ascending further in dimensions, a generalization of the QH effect to 4D was proposed \cite{Zhang:2001}. It has received much attention in theoretical studies \cite{Qi:2008, Edge:2012, Li:2013}, in particular since it might exhibit novel strongly correlated QH phases \cite{Zhang:2001}. This interest was renewed recently due to the unprecedented control and flexibility offered by engineered systems like ultracold atoms and photonics. Indeed, such systems have already been used to study various topological effects \cite{Goldman:2016, Lu:2014}, including in synthetic dimensions \cite{Mancini:2015, Stuhl:2015}, and offer a direct route for realizing 4D physics \cite{Price:2015, Ozawa:2016}.

In the simplest case, a 4D QH system can be composed of two 2D QH systems in orthogonal subspaces (\fig{fig:1}a,~b). In addition to the quantized linear response underlying the 2D QH effect, it exhibits a quantized non-linear 4D Hall response \cite{Kraus:2013}. The latter arises when -- simultaneously with the perturbing electric field $\mathbf{E}$ -- a magnetic perturbation $\mathbf{B}$ is added, coupling the motion in the two 2D QH systems. Due to the 4D symmetry, there are multiple possibilities for the orientation of $\mathbf{E}$ and $\mathbf{B}$. The non-linear response, however, is always characterized by the same 4D topological invariant, the second Chern number. For simplicity, we therefore focus on the geometry depicted in \fig{fig:1}a,~b. In this case, the non-linear response can be understood semiclassically as originating from a Lorentz force created by $\mathbf{B}$ \cite{Price:2016}. The direction of this response is in an orthogonal subspace compared to both perturbing fields and it can thus only occur in four or more dimensions. Therefore, such a response has never been observed in any physical system. Note that a measurement of the second Chern number has recently been reported in an artificially generated parameter space of a four-level system \cite{Sugawa:2016}. 

\invisiblesection{Topological Charge Pumping}
\looseness -1 Topological charge pumps exhibit topological transport properties similar to higher-dimensional QH systems and thus provide a way to probe 4D QH physics in lower-dimensional dynamical systems. The first example for such a topological charge pump is the 1D Thouless pump \cite{Thouless:1983}, where a quantized particle transport is generated by varying the system's parameters in an adiabatic and periodic way. This modulation can be parametrized by a pump parameter and at each point in the cycle, the 1D system constitutes a Fourier component of a 2D QH system \cite{Qi:2008, Kraus:2012b}. The induced motion is thus equivalent to the linear Hall response and characterized by the same 2D topological invariant, the first Chern number. Indeed, the QH effect on a cylinder can be mapped to a 1D charge pump with the threaded flux $\phi_x$ acting as the pump parameter \cite{Laughlin:1981} (\fig{fig:1}a). Building on pioneering experiments in condensed matter systems \cite{Pothier:1992, Switkes:1999}, topological charge pumps have recently been realized in photonic waveguides \cite{Kraus:2012a, Verbin:2015} and with ultracold atoms \cite{Lohse:2016, Nakajima:2016, Lu:2016, Schweizer:2016}.

A dynamical version of the 4D QH effect can be realized by extending the concept of topological charge pumping to 2D \cite{Kraus:2013}. Using dimensional reduction \cite{Qi:2008, Kraus:2012b}, the Fourier components of a 4D QH system can be mapped onto a 2D system. For the geometry in \fig{fig:1}a,~b, the corresponding 2D model is a square superlattice (\fig{fig:1}c and Supplementary Information). It consists of two 1D superlattices along $x$ and $y$, each formed by superimposing two lattices $V\sub{s,\mu} \sin^2 \left( \pi \mu / d\sub{s,\mu} \right) + V\sub{l,\mu} \sin^2 \left( \pi \mu/d\sub{l,\mu} - \varphi_\mu/2 \right)$, $\mu \in \{x,y\}$. Here, $d\sub{s,\mu}$ and $d\sub{l,\mu} > d\sub{s,\mu}$ denote the lattice periods and $V\sub{s,\mu}$ ($V\sub{l,\mu}$) the depth of the short (long) lattice potential, respectively. The position of the long lattices relative to the short ones is determined by the corresponding superlattice phases $\varphi_\mu$.

The phase $\varphi_x$ is chosen as the pump parameter, i.e.~pumping is performed by moving the long lattice along $x$. This is equivalent to threading $\phi_x$ in the 4D model and therefore leads to a quantized motion along $x$ -- the linear response (\fig{fig:1}c, d). The magnetic perturbation $B_{xw}$ corresponds to a transverse superlattice phase $\varphi_y$ that depends linearly on $x$. This can be realized by tilting the long $y$-lattice relative to the short one by an angle $\theta \ll 1$ in the $xy$-plane (\fig{fig:1}c). Then $\varphi_y (x) = \varphi_y^{(0)} + 2 \pi \theta \, x/d\sub{l,y}$ to first order in $ \theta$, thereby coupling the two orthogonal axes. When $\varphi_x$ is varied, the motion along $x$ thus changes $\varphi_y$ and -- analogous to the Lorentz force in 4D -- induces a quantized non-linear response along $y$ \cite{Kraus:2013} (\fig{fig:1}d).

For a uniformly populated band in an infinite system, the change in the centre-of-mass (COM) position during one cycle $\varphi_x = 0 \rightarrow 2\pi$ is
\begin{equation}
\label{eq:1}
\nu_1^x \, d\sub{l,x} \, \mathbf{e}_x + \nu_2 \, \theta \, d\sub{l,x} \, \mathbf{e}_y
\end{equation}
with $\mathbf{e}_x$ ($\mathbf{e}_y$) the unit vector along $x$ ($y$) (see Supplementary Information). The first term describes the quantized linear response in the $x$-direction, proportional to the pump's first Chern number $\nu_1^x$. Here, $\nu_1^x$ is defined using a generalized 2D Brillouin zone spanned by the quasimomentum $k_x$ and $\varphi_x$. It is obtained by integrating the Berry curvature $\Omega^{x} (k_x, \varphi_x) = i \left( \braket{\partial_{\varphi_x} u}{\partial_{k_x} u} - \braket{\partial_{k_x} u}{\partial_{\varphi_x} u}\right)$ over the entire Brillouin zone, where $\ket{u (k_x, \varphi_x)}$ denotes the eigenstate of a given
\clearpage 
\noindent non-degenerate band at $k_x$ and $\varphi_x$. As $\nu_1^x$ can only take integer values, the motion is quantized \cite{Lohse:2016}. The second term is the non-linear response in the $y$-direction. It is quantified by a 4D integer topological invariant, the pump's second Chern number $\nu_2 = 1/(4 \pi^2) \oint\sub{BZ} \Omega^{x} \Omega^{y} \mathrm{d}k_x \mathrm{d}k_y \mathrm{d}\varphi_x \mathrm{d}\varphi_y$, where $\mathrm{BZ}$ indicates the generalized 4D Brillouin zone (\fig{fig:1}e). Therefore, the non-linear response is quantized as well and has intrinsic 4D symmetries resulting from the higher-dimensional non-commutative geometry.

\invisiblesection{Experimental Implementation}
We implement a 2D topological charge pump with bosonic $^{87}$Rb atoms forming a Mott insulator in isolated planes of a 3D optical lattice with superlattices along $x$ and $y$ with $d\sub{s} \equiv d\sub{s,x} = d\sub{s,y}$ and $d\sub{l} \equiv d\sub{l,x} = d\sub{l,y} = 2 d\sub{s}$ (see Methods). This creates double well potentials along both $x$ and $y$ (\fig{fig:1}c). In the tight-binding limit, this realizes a 2D Rice-Mele model \cite{Rice:1982} in each plane, where the on-site energies of the lattice sites are modulated alternatingly by $(-1)^{m_x} \Delta_x (\varphi_x)/2 + (-1)^{m_y} \Delta_y (\varphi_y)/2 $. Here, $m_x$ ($m_y$) denotes the $m_x$-th ($m_y$-th) lattice site along $x$ ($y$). The long lattices also lead to dimerized tunnel couplings between neighbouring sites, $J_x(\varphi_x) + (-1)^{m_x} \delta J_x (\varphi_x)/2$ and $J_y(\varphi_y) + (-1)^{m_y} \delta J_y (\varphi_y)/2$. The corresponding unit cell is a four-site plaquette and the lowest band in the short lattices splits into four subbands. 

In the experiment, we study the non-linear bulk response of the lowest subband, for which $\nu_2 = +1$ is expected for $d\sub{l} = 2 d\sub{s}$. As the initial state, a Mott insulator with quarter filling is prepared at $\varphi_x = 0$, where one atom is localized in the ground state of each unit cell, leading to a uniform occupation of the lowest subband (see Methods). The pumping is performed along $x$ by adiabatically varying $\varphi_x$ and we examine the resulting motion of the atoms.  We locally probe the system's transport properties by using a small atom cloud extending over approximately 20 sites in the $x$-direction. In this case, the variation of $\Omega^y (\varphi_y)$ over the cloud size is negligible and the $y$-displacement per cycle is given by $\overline{\Omega} (\varphi_y^{(0)})\, \theta \, d\sub{l} \, \mathbf{e}_y$ with $ \overline{\Omega} = 1/(2\pi) \oint \Omega^{x}\Omega^{y} \mathrm{d}k_x \mathrm{d}k_y \mathrm{d}\varphi_x$ (see Supplementary Information). From this local response, the global quantized non-linear response of an infinite system can be reconstructed by sampling all possible values of $\varphi_y^{(0)}$, thereby integrating over the entire 4D generalized Brillouin zone.

\begin{figure}[t!]
\includegraphics[width=\linewidth]{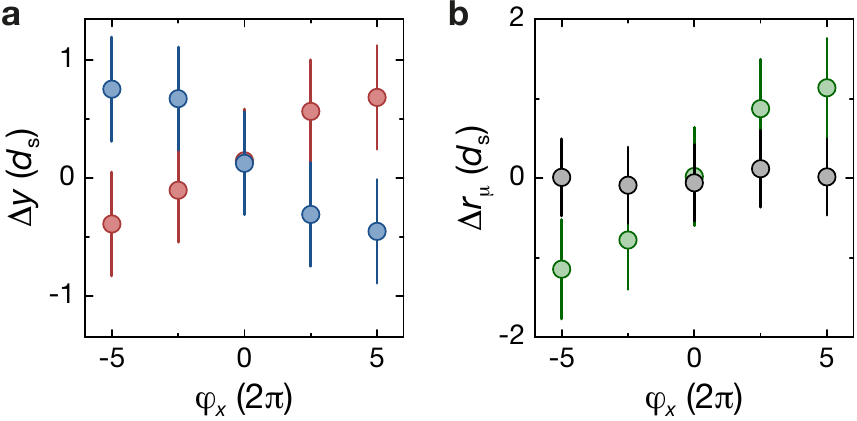}
\caption{4D-type non-linear centre-of-mass (COM) response. \textbf{(a)} COM of the atom cloud along $y$ versus number of pump cycles along $x$ measured for two different angles, $\theta_1 = 0.78(2)\,\mathrm{mrad}$ (red) and $\theta_2 = -0.85(2)\,\mathrm{mrad}$ (blue) with $\varphi_y^{(0)} = 0.500(5)\pi$. When pumping along $x$, the cloud moves in the perpendicular $y$-direction with the sign depending on the pumping direction and the sign of $\theta$. $\Delta y$ is the differential displacement for $V\sub{s,x} = 7.0(2) E\sub{r,s}$, $V\sub{s,y} = 17.0(5) E\sub{r,s}$, $V\sub{l,x} = 20.0(6) E\sub{r,l}$ and $V\sub{l,y} = 80(3) E\sub{r,l}$ compared to a reference sequence with $V\sub{s,y} = 40(1) E\sub{r,s}$ and $V\sub{l,y} = 0 E\sub{r,l}$ (see Methods). Here, $E\sub{r,i} = h^2/(8m\sub{a} d\sub{\mathrm{i}}^2)$, $\mathrm{i} \in\{\mathrm{s,l}\}$, denotes the corresponding recoil energy with $m\sub{a}$ being the mass of an atom. Each point is averaged 100 times and the error bar takes into account the error of the mean as well as a systematic uncertainty of $\pm0.3d\sub{s}$. 
 \textbf{(b)} Difference of the COM drift between $\theta_1$ and $\theta_2$ for the $x$- (grey) and $y$-direction (green), $\Delta r_{\mu} = \Delta \mu(\theta_1) - \Delta \mu(\theta_2)$ with $\mu \in \{x,y\}$. The direction of the non-linear response reverses when changing the sign of $\theta$ whereas the linear response is independent of $\theta$.  Data points are calculated from the measurements in (a) (see Methods).
\label{fig:2}}
\end{figure}

\looseness=-1 To probe the motion of the cloud, we measure its COM position as a function of $\varphi_x$. As the non-linear response results from two weak perturbations, the displacement per cycle is typically only a fraction of $d\sub{l}$. It is thus too small to be resolved experimentally, where the number of cycles is limited due to heating. For suitable lattice parameters, however, where $\overline{\Omega}$ is strongly peaked at $\varphi_y^{(0)} = (l + 1/2) \pi, l \in \mathbb{Z}$ (c.f.~\fig{fig:1}e), signatures of the 4D-like non-linear drift can be seen at $\varphi_y^{(0)} = \pi/2$ (\fig{fig:2}). 

\begin{figure*}[t!]
\includegraphics[width=0.997\textwidth]{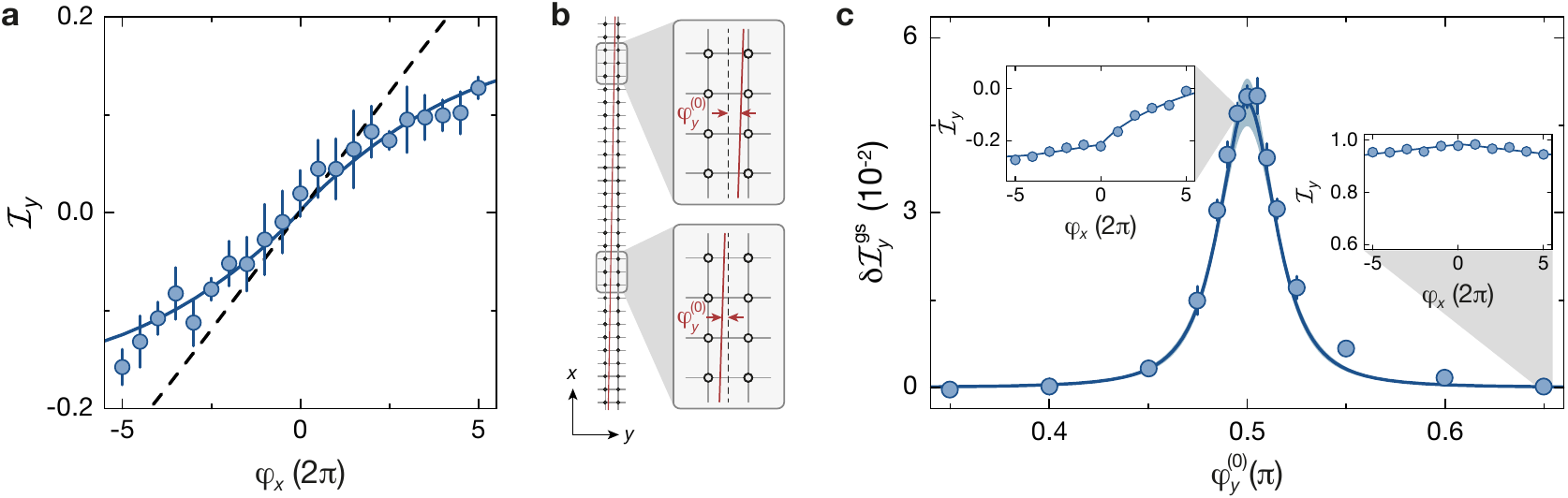}
\caption{Measurement of the second Chern number by local probing of the non-linear response for $\theta = 0.54(3)\,\mathrm{mrad}$. \textbf{(a)}~Double well imbalance $\mathcal{I}_y$ versus number of pump cycles in the $x$-direction at $\varphi_y^{(0)} = 0.500(5)\pi$, $V\sub{s,x} = V\sub{s,y} = 7.0(2) E\sub{r,s}$ and $V\sub{l,x} = V\sub{l,y} = 20.0(6) E\sub{r,l}$. The data points are the average of 14 measurements for the point at $\varphi_x = 0$ and 7 measurements for all others; the error is the error of the mean. The dashed line shows the response of an ideal system; the solid line includes corrections for the finite pumping efficiency along $x$ as well as the creation of doubly-occupied sites and band excitations along $y$ and both curves are shifted by a constant offset $\mathcal{I}_0$ (see Methods). For simplicity, the theory curves assume a homogeneous Berry curvature $\Omega^x = d\sub{l,x}  \nu_1^x/(2\pi)$, neglecting the variation of $\Omega^x$ during a pump cycle.
\textbf{(b)}~The response of an infinite system can be reconstructed with a small atom cloud by repeating the measurement from (a) for different values of $\varphi_y^{(0)}$. A single measurement locally probes the response at the cloud's position (grey frames on the left). Changing $\varphi_y^{(0)}$ is equivalent to sampling a different position in the lattice (insets on the right). Note that the tilt of the long $y$-lattice is greatly exaggerated compared to the angle used in the experiment.
\textbf{(c)}~Change of the double well imbalance per cycle for the lowest band, $\delta \mathcal{I}_y^\mathrm{gs}$, as a function of $\varphi_y^{(0)}$. It is determined by the integrated Berry curvature $\overline{\Omega} (\varphi_y^{(0)})$ and thus exhibits a pronounced peak around $\varphi_y^{(0)} = \pi/2$ (see \fig{fig:1}e and Supplementary Information). The slope $\delta \mathcal{I}_y^\mathrm{gs}$ is extracted from a fit to the measured imbalance $\mathcal{I}_y (\varphi_x)$ (see Methods) and the solid line is the theoretically expected slope. Error bars show the fit error and the blue-shaded region indicates the uncertainty in the slope resulting from the errors of $\theta$ and the lattice depths. The insets show two additional examples of individual measurements of $\mathcal{I}_y (\varphi_x)$ as in (a). 
\label{fig:3}}
\end{figure*}

To quantify this non-linear response, we instead use site-resolved band mapping, which measures the atom number on even ($N_e$) and odd sites ($N_o$) along $y$. This allows for an accurate determination of the average population imbalance of the double wells in the $y$-direction, $\mathcal{I}_y = (N_o - N_e)/(N_o + N_e)$. If no transitions between neighbouring unit cells along $y$ occur, $\mathcal{I}_y$ is directly related to the COM motion (see Methods). An example for a measurement of $\mathcal{I}_y (\varphi_x)$ is shown in \fig{fig:3}a. The measured non-linear response is smaller than expected for an ideal system. The deviation can be attributed to the appearance of doubly-occupied plaquettes and band excitations along $y$ during the pumping as well as a finite pumping efficiency along $x$ (see Methods). The latter is determined by the fraction of atoms remaining in the lowest subband during half a cycle and amounts to 98.6(2)\%. Taking these imperfections into account, we find excellent agreement between the experimental data and the expected imbalance (\fig{fig:3}a). 

By fitting our model for the double well imbalance to the data (see Methods), we can extract the change of the population imbalance for ground state atoms during one cycle, $\delta \mathcal{I}_y^\mathrm{gs} = \mathcal{I}_y^\mathrm{gs} (\varphi_x = 2\pi) - \mathcal{I}_y^\mathrm{gs} (\varphi_x = 0)$. This slope is determined by $\overline{\Omega}$ and thus characterizes the transport properties of the lowest band. To reconstruct the quantized response of an infinite system and thereby obtain $\nu_2$, the measurement of $\mathcal{I}_y(\varphi_x)$ is repeated for different values of $\varphi_y^{(0)}$. This corresponds to using the small atom cloud as a local probe at different positions along $x$ (\fig{fig:3}b). The experimentally determined slope of the non-linear response for the lowest band agrees very well with the one expected in an ideal system (\fig{fig:3}c).

An experimental value for the second Chern number of the lowest subband, $\nu_2^{\mathrm{exp}}$, can be obtained by averaging $\delta \mathcal{I}_y^\mathrm{gs}$ over $\varphi_y^{(0)} \in [0,2\pi[$. For this, the ideal slope is fitted to the measured one by scaling it with a global amplitude, $\left(\nu_2^{\mathrm{exp}}/\nu_2\right) \delta \mathcal{I}_y^\mathrm{gs} (\varphi_y^{(0)})$. For symmetry reasons, it is sufficient to restrict $\varphi_y^{(0)}$ to $[0,\pi[$ for $d\sub{l} = 2 d\sub{s}$ \cite{Marra:2015}. In this interval, the non-linear response has significant contributions only in the vicinity of $\varphi_y^{(0)} = \pi/2$. Fitting the data of \fig{fig:3}c yields $\nu_2^{\mathrm{exp}} = 1.07(8)$, in agreement with the expected value $\nu_2 =  +1$ for $d\sub{l} = 2 d\sub{s}$. The error takes into account the fit error and the uncertainties in the lattice depths and $\theta$. Note that while this approach requires prior knowledge of the response of an ideal system, $\nu_2^{\mathrm{exp}}$ can also be obtained directly from the measurements of $\mathcal{I}_y (\varphi_x)$ without any additional input, which gives $\nu_2^{\mathrm{exp}} = 0.9(2)$ (see Supplementary Information).

\begin{figure}[t!]
\includegraphics[width = \linewidth]{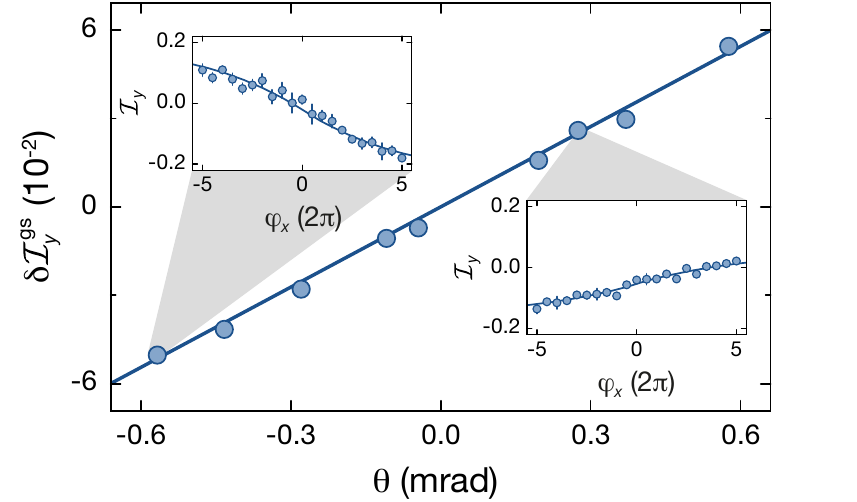}
\caption{Non-linear response versus tilt angle $\theta$ at $\varphi_y^{(0)} = 0.500(5)\pi$. The slope is determined by measuring the double well imbalance when pumping along $x$ as described in \fig{fig:3} and using the same lattice depths. The solid line shows the slope expected for an ideal system. The fit errors for $\delta \mathcal{I}_y^\mathrm{gs}$ are smaller than the size of the data points and the insets show two examples for the measurement of $\mathcal{I}_y (\varphi_x)$ as in \fig{fig:3}a.
\label{fig:4}}
\end{figure}

In the 4D QH system, the non-linear response scales linearly with the magnetic perturbation. The same dependence is thus expected for the non-linear response of our 2D charge pump with respect to $\theta$. We verify this by measuring the peak slope $\delta \mathcal{I}_y^\mathrm{gs}$ at $\varphi_y^{(0)} = \pi/2$ versus $\theta$ (\fig{fig:4}). This provides another way to obtain the second Chern number by determining the slope of $\delta \mathcal{I}_y^\mathrm{gs}(\theta)$ (see Methods). A linear fit gives $\nu_2^{\mathrm{exp}} = 1.01(8)$, where the error is determined as described above. Furthermore, we confirm that the peak slope at fixed $\theta$ scales with the depth of the short $y$-lattice $V\sub{s,y}$ as expected (see Supplementary Information). In particular, the direction of the non-linear response is independent of $V\sub{s,y}$, indicating its robustness against perturbations of the system.

In conclusion, we presented the first measurement of the 4D second Chern number in a bulk system. This first realization of a 2D topological charge pump opens the route to studying higher-dimensional QH physics in experimentally accessible setups. Extending this work, additional density-type non-linear Hall responses can be measured which are implied by the intrinsic 4D symmetry of a 2D charge pump \cite{Kraus:2013}. They manifest differently in the COM drift, but are characterized by the same topological invariant \cite{Price:2016}.  Including interactions might allow for the detection of intriguing fractional phases originating in the 4D fractional QH effect \cite{Zhang:2001}, similar to previous proposals for 1D charge pumps \cite{Zhu:2013}. Furthermore, a QH system with four extended dimensions might be realized with cold atoms \cite{Price:2015} using recently demonstrated techniques for creating synthetic dimensions \cite{Mancini:2015, Stuhl:2015}.\\

\invisiblesection{Notes and Acknowledgments}
\textit{Note:} Simultaneously with this work, complementary results on topological edge states in 2D photonic pumps have been obtained [O. Zilberberg et al., to be published]. \\

We acknowledge insightful discussions with M. Aidelsburger and I. Carusotto. This work was funded by the European Commission (UQUAM, SIQS), the Deutsche Forschungsgemeinschaft (DIP, FOR2414) and the Nanosystems Initiative Munich. M.~L.~was additionally supported by the Elitenetzwerk Bayern (ExQM), H.~M.~P.~by the European Commission (FET Proactive grant AQuS and Marie Sk\l{}odowska--Curie Action, Grant No. 656093 ``SynOptic'') and the Autonomous Province of Trento (SiQuro) and O.~Z.~by the Swiss National Science Foundation.

\bibliographystyle{bosons}

\input{2D_Pumping_Main_Bibliography.bbl}
\newpage
\section*{Methods}

\small
{\bf Lattice configuration.}
All experiments are performed in a mutually orthogonal retro-reflected 3D optical lattice consisting of superlattices along $x$ and $y$ and a simple lattice in the $z$-direction. Each superlattice is created by superimposing two standing waves, a short lattice with wavelength $\lambda\sub{s} = 767$\,nm and a long lattice with $\lambda\sub{l} =2\lambda\sub{s}$. The vertical lattice along $z$ is formed by a standing wave with $\lambda_z = 844$\,nm.

{\bf Tight-binding Hamiltonian of the 2D superlattice.}
\looseness -1 In the tight-binding limit, a 2D superlattice with $d\sub{l} = 2 d\sub{s}$ is described by the 2D Rice-Mele Hamiltonian
\begin{align}
& \hat{H}_{2D}(\varphi_x, \varphi_y) = \nonumber \\
&- \sum_{m_x,m_y} \left[J_x(\varphi_x) + (-1)^{m_x} \delta J_x(\varphi_x)/2 \right] \hat{a}^{\dagger}_{m_x+1,m_y} \hat{a}_{m_x,m_y} + \mathrm{h.c.} \nonumber \\ 
&- \sum_{m_x,m_y} \left[J_y(\varphi_y) + (-1)^{m_y}  \delta J_y(\varphi_y)/2 \right] \hat{a}^{\dagger}_{m_x,m_y+1} \hat{a}_{m_x,m_y} + \mathrm{h.c.} \nonumber \\
&+ \sum_{m_x,m_y} \frac{1}{2} \left[(-1)^{m_x} \Delta_x (\varphi_x) + (-1)^{m_y} \Delta_y(\varphi_y)\right] \hat{a}^{\dagger}_{m_x,m_y} \hat{a}_{m_x,m_y} \nonumber
\end{align}
with $\hat{a}_{m_x,m_y}^{\dagger}$ $(\hat{a}_{m_x,m_y})$ being the creation (annihilation) operator acting on the ($m_x$,$m_y$)-th site in the $xy$-plane. The first (second) term describes the hopping between neigbouring sites along the $x$-axis ($y$-axis) and the last term contains the on-site potential of each lattice site. The long lattices lead to a modulation of the on-site energies, $(-1)^{m_\mu} \Delta_\mu / 2$, and the tunnelling matrix elements, $J_\mu + (-1)^{m_\mu} \delta J_\mu/2$, with $\mu \in \{x,y\}$, which depend on the respective superlattice phase $\varphi_\mu$.

{\bf Initial state preparation for band-mapping measurements.}
For all sequences, a Mott insulator with quarter filling consisting of about 5000 $^{87}$Rb atoms is prepared in the lowest subband of the 2D superlattice. To this end, a Bose-Einstein condensate is loaded from a crossed dipole trap into the lattice by first ramping up the blue-detuned short lattices along $x$ and $y$ to $3.0(1) E\sub{r,s}$ during 50\,ms to lower the initial density of the atom cloud. Then these lattices are switched off again within 50\,ms while at the same time the vertical lattice as well as both long lattices are increased to $30(1) E\sub{r,z}$ and $30(1) E\sub{r,l}$, respectively, with $\varphi_x = 0.000(5) \pi$ and $\varphi_y = \varphi_y^{(0)}$. Subsequently, doubly-occupied lattice sites are converted to singly-occupied ones (see Supplementary Information), creating a Mott insulator with unit filling and a negligible fraction of doublons. Afterwards, each lattice site is split to a four-site plaquette by ramping up the short lattices along $x$ and $y$ to their final depth of $7.0(2) E\sub{r,s}$ and decreasing the long lattices to  $20.0(6) E\sub{r,l}$ within 5\,ms.

{\bf Sequence for pumping.}
The superlattice phase can be controlled by slightly changing the frequency of the lasers used for generating the long lattices and thereby moving the relative position between the short and long lattice at the position of the atoms. The pumping along $x$ is performed by slowly changing $\varphi_x$, starting from the staggered configuration at $\varphi_x = 0.000(5) \pi$, where the energy difference between neighbouring sites $|\Delta_x|$ is largest and the tunnel couplings are equal, $\delta J_x = 0$. To minimize non-adiabatic transitions to higher bands, each pump cycle consists of three s-shaped ramps $\varphi_x \in \left[0,0.5\pi\right], \left[0.5\pi, 1.5 \pi \right]$ and $\left[1.5\pi,2\pi \right]$. This reduces the ramp speed in the vicinity of the symmetric double well configuration ($\Delta_x = 0$) at $\varphi_x = (l + 1/2)\pi$, $l \in \mathbb{Z}$, where the gap to the first excited band is smallest. The duration of the $\pi/2$ ramps is 7\,ms and 14\,ms for the ramp by $\pi$. Due to the limited tuning range of a single laser, a second laser is required for implementing multiple pump cycles, which is set to a constant phase of $\varphi_x = 0.000(5) \pi$. At the end of each cycle, an instantaneous switch from the primary laser to the second one is made and within 5\,ms the frequency of the former is ramped back to its initial value --- corresponding to an identical lattice configuration. After switching back to the first laser, the next cycle continues as described above. We checked experimentally that this handover between the two lasers does not create any measurable band excitations.

{\bf Measurement of the in-situ position.}
To determine the non-linear COM displacement along $y$, a double-differential measurement is conducted to minimize the effect of shot-to-shot fluctuations of the atom position. In order to do this, the COM position is measured before ($y\sub{i}$) and after the pumping ($y\sub{f}$) and compared to a reference sequence  ($y\sub{i}^{(0)}$, $y\sub{f}^{(0)}$). In the latter, the pumping is performed with only the short lattice along $y$ (at $V\sub{s,y} = 40(1) E\sub{r,s}$) and therefore the non-linear response is zero. The initial position is obtained during the doublon removal sequence, where the atoms are initially prepared in the $(F=1, m_F = 0)$ hyperfine state and one atom from each doubly-occupied site is transferred to $(F=2, m_F = -2)$ using microwave-dressed spin-changing collisions (see Supplementary Information). Here, $F$ denotes the total angular momentum of the atoms. In addition, we transfer 50\% of the atoms on singly-occupied sites to the $F = 2$ manifold as well by applying a microwave $\pi$-pulse resonant on the $(F=1, m_F = 0) \rightarrow (F=2, m_F = 0)$ transition. The $F = 2$ atoms thus have the same density distribution as the remaining $F = 1$ atoms and are imaged prior to the push-out pulse, which removes them from the lattice. The motion of the atoms due to the non-linear response is then given by $\Delta y = (y\sub{f} - y\sub{i}) - (y\sub{f}^{(0)} - y\sub{i}^{(0)})$. The difference of the COM displacement along $y$ between $\theta_1$ and $\theta_2$ is defined as $\Delta r_y = \Delta y (\theta_1) - \Delta y (\theta_2)$. For the $x$-direction, it is obtained directly from $\Delta x = (x\sub{f} - x\sub{i}) - \delta \overline{x}$ without comparing it to the reference sequence. Here, $\delta \overline{x}$ is the average displacement of all data points for a given angle, accounting for a small constant offset between the measured initial and final positions.

{\bf Relation between centre-of-mass position and double well imbalance.}
If there are no inter-double-well transitions along $y$, the change in the double well imbalance $\delta \mathcal{I}_y  = \mathcal{I}_y (\varphi_x) - \mathcal{I}_y (\varphi_x = 0)$ can be directly related to the COM motion along $y$. The COM position in the $y$-direction is given by $y\sub{COM} = d\sub{l}/N \, \sum_{ij} \left[ (j-1/4) N_{e,ij} + (j+1/4) N_{o,ij}\right]$, where the sum is over all unit cells, $N_{e,ij}$ ($N_{o,ij}$) is the occupation of the even (odd) sites along $y$ in the $(i,j)$-th unit cell and $N$ is the total atom number. Expressing this in terms of the total number of atoms on even and odd sites, $N_e = \sum_{ij} N_{e,ij}$ and $N_o = \sum_{ij} N_{o,ij}$, and assuming that there are no transitions between neighbouring unit cells along $y$, i.e.~$\sum_i N_{e,ij} + N_{o,ij}$ remains constant, the change in the COM position can be written as $\delta y\sub{COM}  = y\sub{COM} (\varphi_x) - y\sub{COM} (\varphi_x = 0) = d\sub{l} \, \delta \mathcal{I}_y/4$. Note that this derivation implicitly assumes that the COM of the maximally-localized Wannier functions on the lattice sites along $y$ is independent of $\varphi_y$, which is a valid approximation deep in the tight-binding regime. Otherwise, the proportionality factor $d\sub{l}/2$ has to be replaced by the distance between the COM of the Wannier functions on the even and odd site of a double well.

{\bf Model for double well imbalance including experimental imperfections.}
To determine the slope of the non-linear response from the band mapping data, we use a simple model that takes into account band excitations and double occupation of plaquettes and the experimental pumping efficiency of the linear response. The average double well imbalance $\mathcal{I}_y (\varphi_x)$ can be written as 
\begin{equation*}
\label{eq:M1}
\mathcal{I}_y (\varphi_x) = n\sub{gs} \mathcal{I}_y^\mathrm{gs} (\varphi_y) + n\sub{exc} \mathcal{I}_y^\mathrm{exc} (\varphi_y) + n\sub{2} \mathcal{I}_y^\mathrm{2,gs} (\varphi_y)
\end{equation*}
where $n\sub{gs}$ ($n\sub{exc}$) is the fraction of atoms on singly-occupied plaquettes in the ground (first excited) state along $y$ and $n\sub{2}$ is the fraction of atoms on doubly-occupied plaquettes, which we assume to be in the ground state. These quantities can be determined experimentally at each point in the pumping sequence. $\mathcal{I}_y^\mathrm{gs}$, $\mathcal{I}_y^\mathrm{exc}$ and $\mathcal{I}_y^\mathrm{2,gs}$ denote the imbalance of the corresponding state, which is determined by the local phase of the $y$-superlattice at the position of the cloud along $x$, $\varphi_y (x\sub{COM})$, and can be calculated using the respective double well Hamiltonian (see Supplementary Information). The COM position in turn depends on the pump parameter $\varphi_x$ and includes corrections for the finite pumping efficiency, $x\sub{COM} (\varphi_x) = \mathrm{sgn}(\varphi_x) \sum_{i=1}^{|\varphi_x|/\pi} \left( 2 \beta_0 \beta^i - \beta \right)$ for $\varphi_x/\pi \in \mathbb{Z}$. Here, $\beta_0 = 0.980(4)$ is the initial ground state occupation along $x$ and $\beta = 0.986(2)$ is the pumping efficiency, given by the fraction of atoms that are transferred by one lattice site along $x$ during each half of a pump cycle. The main contributions limiting the pumping efficiency are band excitations in the pumping direction as well as  non-adiabatic transitions between neighbouring double wells induced by the external harmonic confinement.

{\bf Fit function for non-linear response.}
Based on the above model, the experimental data is fitted with the function $\mathcal{I}_y (\varphi_x) + \mathcal{I}_0$ with $\varphi_y \rightarrow \varphi_y\supscr{exp} = \varphi_y^{(0)} + \alpha \, (\varphi_y - \varphi_y^{(0)})$. The two fit parameters are the prefactor $\alpha$, which describes the change of the superlattice phase along $y$ with $\varphi_x$ compared to the ideal case $\varphi_y\supscr{exp} = \varphi_y$, and an overall offset $\mathcal{I}_0$. The transport properties of the lowest band are encoded in the slope of the ground state imbalance at $\varphi_x = 0$. Knowing $\alpha$, it can be related to the ideal slope via 
$$ \frac{\partial \mathcal{I}_y^\mathrm{gs}(\varphi_y\supscr{exp}) }{ \partial \varphi_x} = \frac{\partial \mathcal{I}_y^\mathrm{gs}(\varphi_y\supscr{exp})}{\partial \varphi_y\supscr{exp}} \, \frac{\partial \varphi_y\supscr{exp}}{\partial \varphi_x}  = \alpha \, \frac{\partial \mathcal{I}_y^\mathrm{gs}(\varphi_y) }{ \partial \varphi_x}$$
Per cycle, this gives a change of the population imbalance for ground state atoms of 
$$\delta \mathcal{I}_y^{\mathrm{gs}}  = \alpha \left[  \mathcal{I}_y^\mathrm{gs}(\varphi_y)\Big|_{\varphi_x = 2\pi} - \mathcal{I}_y^\mathrm{gs}(\varphi_y)\Big|_{\varphi_x = 0} \right]$$

{\bf Determination of the second Chern number from scaling of the non-linear response with $\theta$.}
The COM displacement per cycle along $y$ for an infinite system, $\delta y\sub{COM} = \nu_2 \theta d\sub{l,x}$, scales linearly with the perturbing angle $\theta$. The second Chern number can thus be extracted from the slope of $\delta y\sub{COM} (\theta)$. Having confirmed that the measured shape of $\delta \mathcal{I}_y^\mathrm{gs} (\varphi_y^{(0)})$ is the same as expected theoretically, the response of an infinite system at a given angle $\theta$ can be inferred from a single measurement of $\delta \mathcal{I}_y^\mathrm{gs}$ at a fixed $\varphi_y^{(0)}$. This holds for all angles since the shape of $\overline{\Omega} (\varphi_y^{(0)})$ is independent of $\theta$. To obtain $\nu_2$, it is therefore sufficient to determine the slope of $\delta \mathcal{I}_y^\mathrm{gs}(\theta)$ at a constant $\varphi_y^{(0)}$.

\clearpage

\onecolumngrid
\clearpage
\begin{center}
\noindent\textbf{Supplementary Information for:}
\\\bigskip
\noindent\textbf{\large{Exploring 4D Quantum Hall Physics with a 2D Topological Charge Pump}}
\\\bigskip
Michael Lohse$^{1,2}$, Christian Schweizer$^{1,2}$, Hannah M. Price$^{3}$, Oded Zilberberg$^{4}$ \& Immanuel Bloch$^{1,2}$
\\\vspace{0.1cm}
\small{$^{1}$\emph{Fakult\"at f\"ur Physik, Ludwig-Maximilians-Universit\"at, Schellingstra\ss e 4, 80799 M\"unchen, Germany}}\\
\small{$^{2}$\emph{Max-Planck-Institut f\"ur Quantenoptik, Hans-Kopfermann-Stra\ss e 1, 85748 Garching, Germany}}\\
\small{$^{3}$\emph{INO-CNR BEC Center \& Dipartimento di Fisica, Universit\`a di Trento, Via Sommarive 14, 38123 Povo, Italy}}\\
\small{$^{4}$\emph{Institut f\"ur Theoretische Physik, ETH Z\"{u}rich, Wolfgang-Pauli-Stra\ss e 27, 8093 Z\"urich, Switzerland}}
\end{center}
\bigskip
\bigskip
\twocolumngrid
\normalsize


\renewcommand{\thefigure}{S\the\numexpr\arabic{figure}-10\relax}
 \setcounter{figure}{10}
\renewcommand{\theequation}{S.\the\numexpr\arabic{equation}-10\relax}
 \setcounter{equation}{10}
 \renewcommand{\thesection}{S.\Roman{section}}
\setcounter{section}{10}
\renewcommand{\bibnumfmt}[1]{[S#1]}
\renewcommand{\citenumfont}[1]{S#1}

\section{Hall response of the 4D quantum Hall system}
Assuming perfect adiabaticity, the Hall response of the 4D system shown in Fig.~1a,~b can be evaluated from the semiclassical equations of motion for a wave packet centred at position $\mathbf{r}$ and quasimomentum $\mathbf{k}$ \cite{Xiao:2010_SI}: 
\begin{gather}
\label{eq:S1}
\dot{r}^{\mu} = \frac{1}{\hbar} \frac{\partial \mathcal{E}(\mathbf{k})}{\partial k_{\mu}} + \dot{k}_{\nu} \, \Omega^{\nu \mu} (\mathbf{k})\\
\label{eq:S2}
\hbar \dot{k}_{\mu} = q E_\mu + q \dot{r}^\nu B_{\mu \nu}
\end{gather}
Here, $\mathcal{E}(\mathbf{k})$ is the energy of the respective eigenstate at $\mathbf{k}$, $q$ the charge of the particle and the Einstein notation is used for the spatial indices ${\mu, \nu} \in \{w,x,y,z\}$\footnote[1]{Note that the orientation of the axes in Fig. 1a,~b is chosen such that the 4D Levi-Civita symbol is $\varepsilon_{wxyz} = +1$}. The velocity of the wave packet $\mathbf{v} = \dot{\mathbf{r}}$ has two contributions: the group velocity arising from the dispersion of the band and the anomalous velocity due to the non-zero Berry curvature $\Omega^{\nu \mu } (\mathbf{k}) = i \left( \braket{\partial_{k_\nu} u }{\partial_{k_\mu} u} - \braket{\partial_{k_\mu} u }{\partial_{k_\nu} u} \right)$. For a filled or homogeneously populated band, the group velocity term vanishes and with $\mathbf{E} = E_z \mathbf{e}_z$ and $\mathbf{B} = 0$, the linear Hall response is given by the COM velocity
\begin{equation}
\label{eq:S3}
\mathbf{v}\sub{COM}^{(0)} = \frac{q}{h} A\sub{M}^{xz} E_z \nu_1^{zx} \, \mathbf{e}_x
\end{equation}
\looseness=-1 where $A\sub{M}^{xz}$ denotes the size of the magnetic unit cell and $\nu_1^{zx} = 1/(2\pi) \oint\sub{BZ} \Omega^{z x} \, \mathrm{d}^2 k$ the first Chern number of the 2D QH system in the $xz$-plane. The integration is performed over the 2D Brillouin zone spanned by $k_x$ and $k_z$. 

Adding the perturbing magnetic field $B_{xw}$ generates a Lorentz force acting on the moving cloud, $\hbar \dot{\mathbf{k}} = q E_z \, \mathbf{e}_z - q v_x^{(0)} B_{xw} \, \mathbf{e}_w$ \cite{Price:2015_SI}. Note that this additional force can alternatively be interpreted as arising from a Hall voltage in the $w$-direction that is created by the current along $x$ in the presence of $B_{xw}$. This force in turn induces an additional anomalous velocity along $y$, giving rise to the non-linear Hall response. The resulting average velocity is then 
\begin{equation}
\label{eq:S4}
\mathbf{v}\sub{COM} = \frac{q}{h} A\sub{M}^{xz} E_z \nu_1^{zx} \, \mathbf{e}_x - \left(\frac{q}{h}\right)^2 A_M \, E_z B_{xw} \nu_2 \, \mathbf{e}_y
\end{equation}
with $A_M$ being the size of the 4D magnetic unit cell. The second Chern number is given by $\nu_2 = 1/(4 \pi^2) \oint\sub{BZ} \Omega^{xw} \Omega^{zy} + \Omega^{xy} \Omega^{wz} + \Omega^{zx} \Omega^{wy} \mathrm{d}^4 k$, where $\mathrm{BZ}$ denotes the 4D Brillouin zone.

\section{Mapping of a 2D Topological Charge Pump to a 4D Quantum Hall System}
The Hamiltonian of a 2D topological charge pump for a given set of parameters $\{\varphi_x, \varphi_y\}$ can be interpreted as a Fourier component of a higher-dimensional quantum Hall system. Using the approach of dimensional extension \cite{Kraus:2012b_SI}, a 2D charge pump can be mapped onto a 4D QH system, whose Fourier components are sequentially sampled during a pump cycle. This is demonstrated in the following for the deep tight-binding limit $V\sub{s,\mu} \gg V\sub{l,\mu}^2/(4 E\sub{r,s})$, $\mu \in \{x,y\}$, where the corresponding 4D system consists of two 2D Harper-Hofstadter-Hatsugai models \cite{Harper:1955_SI, Azbel:1964_SI, Hofstadter:1976_SI, Hatsugai:1990_SI} in the $xz$- and $yw$-plane. A similar analogy can be made in the opposite limit of a vanishing short lattice, $V\sub{s,x} \rightarrow 0$ and $V\sub{s,y} \rightarrow 0$. In this case, each axis of the 2D lattice maps onto the Landau levels of a free particle in an external magnetic field in 2D \cite{Lohse:2016_SI}. For the lowest band, these two limiting cases are topologically equivalent, i.e. they are connected by a smooth crossover without closing the gap to the first excited band. The topological invariants governing the linear and non-linear response are thus independent of the depth of the short lattices.

For non-interacting atoms in the tight-binding limit, the motion in a 2D superlattice is captured by the following Hamiltonian: 
\begin{align}
\label{eq:S5}
&\hat{H}_{2D}(\varphi_x, \varphi_y) = \\&- \sum_{m_x,m_y} \left[J_x(\varphi_x) + \delta J_x^{m_x}(\varphi_x) \right] \hat{a}^{\dagger}_{m_x+1,m_y} \hat{a}_{m_x,m_y} + \mathrm{h.c.} \nonumber \\ 
&- \sum_{m_x,m_y} \left[J_y(\varphi_y) + \delta J_y^{m_y}(\varphi_y) \right] \hat{a}^{\dagger}_{m_x,m_y+1} \hat{a}_{m_x,m_y} + \mathrm{h.c.} \nonumber \\
&+ \sum_{m_x,m_y} \left[\Delta_x^{m_x} (\varphi_x) + \Delta_y^{m_y} (\varphi_y)\right] \hat{a}^{\dagger}_{m_x,m_y} \hat{a}_{m_x,m_y} \nonumber
\end{align}
Here, $\hat{a}_{m_x,m_y}^{\dagger}$ $(\hat{a}_{m_x,m_y})$ are the creation (annihilation) operator acting on the ($m_x$,$m_y$)-th site in the $xy$-plane, $J_\mu + \delta J_\mu^{m_\mu}$ with $\mu \in \{x,y\}$ are the tunneling matrix elements between neighbouring sites along $\mu$ and $\Delta_x^{m_x} + \Delta_y^{m_y}$ is the on-site potential of a given site. In the presence of the long lattices, the tunnel couplings as well as the on-site energies are modulated periodically by $\delta J^{m_{\mu}}_{\mu}$ and $\Delta_x^{m_x} + \Delta_y^{m_y}$, respectively. For the lattice configuration used in the experiment, where $d\sub{l,\mu} = 2 d\sub{s,\mu}$, these modifications can be expressed as $(-1)^{m_\mu} \delta J_{\mu} (\varphi_{\mu})/2$ and $(-1)^{m_\mu} \Delta_{\mu} (\varphi_{\mu})/2$. 

In the deep tight-binding regime, $J_x$ and $J_y$ become independent of the superlattice phases and the modulations can be approximated as 
\begin{gather}
\delta J_x^{m_x} (\varphi_x) = - \frac{\delta J_x^{(0)}}{2} \sin \left( \tilde{\Phi}_{xz} m_x - \varphi_x\right)\\
\delta J_y^{m_y} (\varphi_y) = - \frac{\delta J_y^{(0)}}{2} \sin \left( \tilde{\Phi}_{yw} m_y - \varphi_y\right)\\
\Delta_x^{m_x} (\varphi_x) = \frac{\Delta_x^{(0)}}{2} \sin \left( \tilde{\Phi}_{xz} (m_x-1/2) - \varphi_x\right)\\
\Delta_y^{m_y} (\varphi_y) = \frac{\Delta_y^{(0)}}{2} \sin \left( \tilde{\Phi}_{yw} (m_y-1/2) - \varphi_y\right)
\end{gather}
with $\tilde{\Phi}_{xz} = 2 \pi d\sub{s,x}/d\sub{l,x}$ and $\tilde{\Phi}_{yw} = 2 \pi d\sub{s,y}/d\sub{l,y}$. In this case, $\hat{H}_{2D}$ is equivalent to the generalized 2D Harper model \cite{Harper:1955_SI, Roux:2008_SI} which describes the Fourier components of a 4D lattice model with two uniform magnetic fields. The 4D parent Hamiltonian can be obtained by performing an inverse Fourier transform \cite{Kraus:2013_SI}
\begin{equation}
\label{eq:S10}
\hat{H}_{4D} = \frac{1}{4\pi^2}\int_0^{2\pi} \hat{H}_{2D} (\varphi_x, \varphi_y) \mathrm{d} \varphi_x \mathrm{d} \varphi_y^{(0)}
\end{equation}
with 
\begin{gather}
\hat{a}^{\dagger}_{m_x,m_y} = \sum_{m_z,m_w} e^{i (\varphi_x m_z + \varphi_y^{(0)} m_w)} \hat{a}^{\dagger}_{\mathbf{m}} \\
\hat{a}_{m_x,m_y} = \sum_{m_z,m_w} e^{-i (\varphi_x m_z + \varphi_y^{(0)} m_w)} \hat{a}_{\mathbf{m}}
\end{gather}
where $\mathbf{m} = \{m_x, m_y, m_z, m_w\}$ indicates the position in the 4D lattice. This yields
\begin{equation}
\label{eq:S13}
\hat{H}_{4D} = \hat{H}_{xz} + \hat{H}_{yw} + \hat{H}_{\delta J}
\end{equation}
The first term $\hat{H}_{xz}$ describes a 2D Harper-Hofstadter model \cite{Harper:1955_SI, Azbel:1964_SI, Hofstadter:1976_SI} in the $xz$-plane with a uniform magnetic flux per unit cell $\Phi_{xz} = \Phi_0 \tilde{\Phi}_{xz}/(2\pi) = d\sub{s,x}/d\sub{l,x} \, \Phi_0$ with $\Phi_0$ denoting the magnetic flux quantum:
\begin{align}
\label{eq:S14}
\hat{H}_{xz} = &- \sum_{\mathbf{m}}  J_x \hat{a}^{\dagger}_{\mathbf{m} + \mathbf{e}_x} \hat{a}_{\mathbf{m}} + \mathrm{h.c.} \\
&- \sum_{\mathbf{m}} \frac{\Delta_x^{(0)}}{4} e^{i \left[\tilde{\Phi}_{xz} (m_x - 1/2) + \pi/2 \right]} \hat{a}^{\dagger}_{\mathbf{m} + \mathbf{e}_z} \hat{a}_{\mathbf{m}} + \mathrm{h.c.} \nonumber
\end{align}
Correspondingly, the second term $\hat{H}_{yw}$ is an independent 2D Harper-Hofstadter model in the $yw$-plane with $\Phi_{yw}= d\sub{s,y}/d\sub{l,y} \, \Phi_0$. Due to the position dependence of the transverse superlattice phase $\varphi_y$, it also contains the magnetic perturbation, i.e.~a weak homogeneous magnetic field in the $xw$-plane:
\begin{align}
\label{eq:S15}
\hat{H}_{yw} &= \\ - &\sum_{\mathbf{m}}  J_y \hat{a}^{\dagger}_{\mathbf{m} + \mathbf{e}_y} \hat{a}_{\mathbf{m}} + \mathrm{h.c.} \nonumber \\
- &\sum_{\mathbf{m}} \frac{\Delta_y^{(0)}}{4} e^{i \left[\tilde{\Phi}_{yw} (m_y - 1/2) + \tilde{\Phi}_{xw} m_x + \pi/2\right]} \hat{a}^{\dagger}_{\mathbf{m} + \mathbf{e}_w} \hat{a}_{\mathbf{m}} + \mathrm{h.c.} \nonumber
\end{align}
with $\tilde{\Phi}_{xw} = -2\pi \theta d\sub{s,x}/d\sub{l,y}$. The strength of the perturbing magnetic field is thus given by
\begin{equation}
\label{eq:S16}
B_{xw} = - \frac{\Phi_0}{d\sub{s,w} d\sub{l,y}} \, \theta
\end{equation}
where $d\sub{s,w}$ is the lattice spacing along $w$. For $\delta J_{\mu}^{(0)} \neq 0$, the third contribution $\hat{H}_{\delta J}$ leads to the appearance of additional diagonal tunnel coupling elements in the $xz$- and $yw$-plane with an amplitude of $\delta J_x^{(0)}/4$ and $\delta J_y^{(0)}/4$, respectively. The individual 2D models without the magnetic perturbation $B_{xw}$ then correspond to the Harper-Hofstadter-Hatsugai model \cite{Hatsugai:1990_SI} with a uniform magnetic flux $\Phi_{xz}$ and $\Phi_{yw}$, respectively, i.e.~the same flux as for $\delta J_{\mu}^{(0)} = 0$.

\section{Transport properties of a 2D topological charge pump}

When the pump parameter $\varphi_x$ is changed slowly, a particle that is initially in an eigenstate $\ket{u (k_x, \varphi_x (t=0), k_y, \varphi_y)}$ of the 2D superlattice Hamiltonian $\hat{H}_{2D}$ [\eq{eq:S5}] will adiabatically follow the corresponding instantaneous eigenstate $\ket{u (k_x, \varphi_x (t), k_y, \varphi_y)}$. In absence of a tilt, $\theta = 0$, the particle acquires an anomalous velocity $\Omega^x \partial_t \varphi_x \mathbf{e}_x$ during this evolution, analogous to the linear Hall response in a QH system. In this case, the Berry curvature $\Omega^x$ is defined in a 4D generalized Brillouin zone $(k_x, \varphi_x, k_y, \varphi_y)$:
\begin{equation}
\label{eq:S17}
\Omega^x (k_x, \varphi_x, k_y, \varphi_y) = i \left( \braket{\partial_{\varphi_x} u}{\partial_{k_x} u} - \braket{\partial_{k_x} u}{\partial_{\varphi_x} u}\right)
\end{equation}
For a homogeneously populated band, the COM displacement along $x$ during one cycle, obtained by integrating the average anomalous velocity over one period, can be expressed as an integral of the Berry curvature over the 2D generalized Brillouin zone spanned by $k_x$ and $\varphi_x$. It is thus determined by the pump's first Chern number
\begin{equation}
\label{eq:S18}
\nu_1^x = \frac{1}{2\pi} \oint \Omega^x \, \mathrm{d} k_x \mathrm{d} \varphi_x 
\end{equation}

When a tilt is present, $\theta \neq 0$, this motion along $x$ leads to a change in $\varphi_y$. This induces an additional anomalous velocity in the $y$-direction, giving rise to the non-linear response. Neglecting the contribution from the group velocity (which averages to zero for a homogeneously populated band), we obtain for a given eigenstate:
\begin{equation}
\label{eq:S19}
v_y (k_x, \varphi_x, k_y, \varphi_y) = \Omega^y \partial_t \varphi_y = \frac{2\pi}{d\sub{l,y}} \theta \, \Omega^x \Omega^y \partial_t \varphi_x
\end{equation}
The distribution of $\Omega^x \Omega^y$ in the 4D generalized Brillouin zone is shown in Fig.~1e for the lattice parameters used for the measurements in Fig.~3 and 4 of the main text. It exhibits a pronounced peak around $\varphi_x \in \{\pi/2, 3\pi/2\}$ and $\varphi_y \in \{\pi/2, 3\pi/2\}$.

For a small cloud that homogeneously populates a single band as in the experiment, the variation of $\Omega^x \Omega^y$ over the size of the cloud due to the position dependence of $\varphi_y$ is negligible for $\theta \ll 1$. The average velocity for the non-linear response can then be calculated by averaging \eq{eq:S19} over both quasimomenta $k_x$ and $k_y$. The COM displacement after a complete cycle can be determined by integrating the velocity over one period. We can thus express the change in the COM position per cycle as
\begin{equation}
\label{eq:S20}
\delta y\sub{COM} = \underbrace{\frac{1}{2\pi} \oint \Omega^x \Omega^y \, \mathrm{d} k_x \mathrm{d} k_y \mathrm{d} \varphi_x}_{\overline{\Omega} (\varphi_y)}   \,\theta \, d\sub{l,x}
\end{equation}
If the number of pump cycles is small, the change of $\varphi_y$ as a result of the linear pumping response can be neglected and the non-linear displacement per cycle is very well approximated by $\delta y\sub{COM} \approx \overline{\Omega} (\varphi_y^{(0)}) \, \theta \, d\sub{l,x}$. 

\looseness -1 The response of a large system with size $L_x \gg d\sub{l,y}/\theta$ can be obtained by averaging \eq{eq:S20} over $\varphi_y(x) \in [0,2\pi[$, yielding 
\begin{equation}
\label{eq:S21}
\delta y\sub{COM} = \frac{1}{2\pi} \oint  \overline{\Omega} (\varphi_y) \, \theta \, d\sub{l,x} \, \mathrm{d}\varphi_y = \nu_2 \, \theta \, d\sub{l,x}
\end{equation}
where the second Chern number $\nu_2$ is calculated by integrating $\Omega^x \Omega^y$ over the entire 4D generalized Brillouin zone:
\begin{equation}
\label{eq:S22}
\nu_2 = 1/(4 \pi^2) \oint\sub{BZ} \Omega^{x} \Omega^{y} \mathrm{d}k_x \mathrm{d}k_y \mathrm{d}\varphi_x \mathrm{d}\varphi_y
\end{equation}

\section{Pump Path}
\begin{figure*}[t!]
\includegraphics[width=0.945\linewidth]{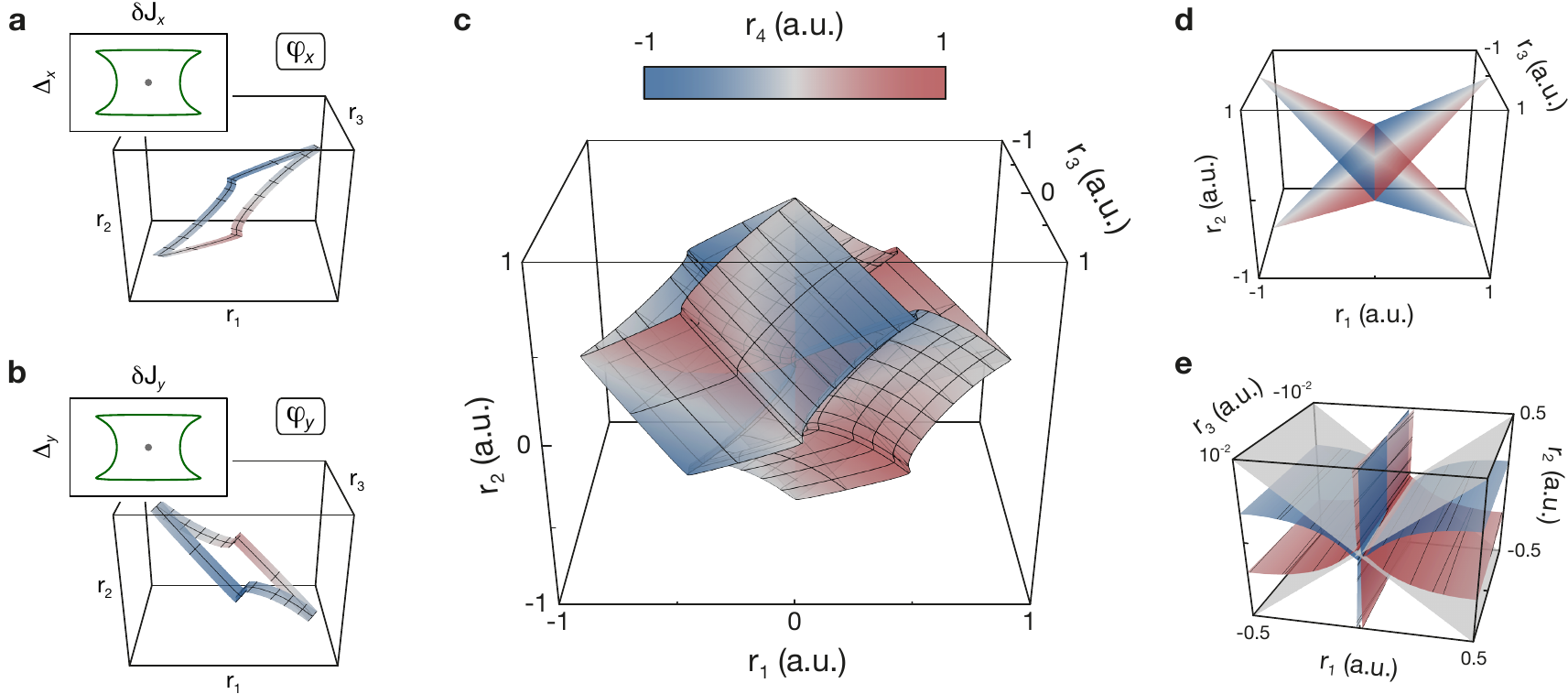}
\caption{ \looseness -1
Pump cycle of the 2D topological charge pump. The 4D tight-binding parameter space ($\delta J_x$, $\Delta_x$, $\delta J_y$, $\Delta_y$) is visualized using the transformation of \eq{eq:S23}. \textbf{(a)} Changing the pump parameter $\varphi_x$ leads to a periodic modulation of $\delta J_x$ and $\Delta_x$ along a closed trajectory as shown in the inset for a full pump cycle $\varphi_x = 0 \rightarrow 2\pi$. This pump path (green) encircles the degeneracy point at the origin (grey), where the gap between the two lowest subbands of the Rice-Mele model closes. The surface in the main plot shows the same trace transformed according to \eq{eq:S23} and with $\varphi_y \in [0.46 \pi, 0.54\pi]$. The spacing of the mesh grid illustrating $\varphi_x$ is $\pi/10$. \textbf{(b)} For a given $\varphi_x$, a large system simultaneously samples all values of $\varphi_y$. This corresponds to a closed path in the $\delta J_y$-$\Delta_y$ parameter space where a singularity occurs at the origin as well (inset). The main plot shows the transformed path for $\varphi_x \in [0.46 \pi, 0.54\pi]$. \textbf{(c)} In a full pump cycle, such a system thus covers a closed surface in the 4D parameter space by translating the path shown in (b) along the trajectory from (a). \textbf{(d)} In the transformed parameter space, the singularities at ($\delta J_x = 0$, $\Delta_x = 0$) and ($\delta J_y = 0$, $\Delta_y = 0$) correspond to two planes that touch at the origin. \textbf{(e)} Cut around $r_3 = 0$ showing both the pump path from (c) (red/blue) as well as the singularities from (d) (grey). While they intersect in the 3D space $(r_1, r_2, r_3)$, the value of $r_4$ is different on both surfaces and the 4D pump path thus fully encloses the degeneracy planes.
\label{fig:S2}}
\end{figure*}

Varying the pump parameter $\varphi_x$ periodically modulates the tight-binding parameters $\delta J_x (\varphi_x)$ and $\Delta_x (\varphi_x)$ describing the superlattice along $x$ [\eq{eq:S5}]. For $d\sub{l} = 2 d\sub{s}$, the modulation of $\delta J_x$ and $\Delta_x$ is out of phase and the system therefore evolves along a closed trajectory in the $\delta J_x$--$\Delta_x$ parameter space (\extfig{fig:S2}a). This pump path encircles the degeneracy point  ($\delta J_x = 0$, $\Delta_x = 0$), where the two lowest subbands of the Rice-Mele model touch. This singularity can be interpreted as the source of the non-zero Berry curvature $\Omega^x$ in the generalized Brillouin zone, which gives rise to the linear pumping response. All pump paths that encircle the degeneracy can be continuously transformed into one another without closing the gap to the first excited subband and are thus topologically equivalent with respect to the linear response, i.e.~the value of $\nu_1^x$ does not change.

Similarly, the tight-binding parameters $\delta J_y$ and $\Delta_y$ depend on the phase of the transverse superlattice $\varphi_y$. For a large cloud, all possible values of $\varphi_y$ and thus $\delta J_y$ and $\Delta_y$ are sampled simultaneously (\extfig{fig:S2}b). During a pump cycle, the system therefore traces out a closed surface in the 4D parameter space of $\delta J_x$, $\Delta_x$, $\delta J_y$ and $\Delta_y$ (\extfig{fig:S2}c). In this parameter space, the two lowest subbands touch in the two planes ($\delta J_x = 0$, $\Delta_x = 0$) and ($\delta J_y = 0$, $\Delta_y = 0$), which intersect in a single point at the origin (\extfig{fig:S2}d). Analogous to the linear response, this degeneracy generates the non-zero Berry curvatures $\Omega^x$ and $\Omega^y$, which cause the non-linear motion in the $y$-direction. Due to the 4D character of the parameter space, the 4D pump path can enclose the degeneracy (\extfig{fig:S2}e). Whenever this is the case, the topology of the cycle does not change and the value of $\nu_2$ remains the same.

To visualize the pump path in the 4D parameter space in \extfig{fig:S2}, we apply the following transformation:
\begin{equation}
\label{eq:S23}
\left( \begin{array}{c} r_1 \\ r_2  \\ r_3 \\ r_4 \end{array} \right) 
=
\frac{1}{4} \left( \begin{array}{cccc} 1 & 1 & -1 & -1 \\  1  & 1 & 1 & 1  \\ 1  & -1  & -1  & 1  \\  1  & -1 & 1 & -1 \end{array} \right) 
\cdot
\left( \begin{array}{c} \delta J_x / \delta J_x^{(0)} \\ \Delta_x / \Delta_x^{(0)} \\ \delta J_y /  \delta J_y^{(0)} \\ \Delta_y / \Delta_y^{(0)} \end{array} \right) 
\end{equation}
where the tight-binding parameters are normalized by their respective maximum values. The degeneracy planes are then given by $r_1 = -r_2$, $r_3 = -r_4$ and $r_1 = r_2$, $r_3 = r_4$, respectively, i.e.~they become perpendicular planes in the $(r_1, r_2, r_3)$-space.

\section{Calculation of the double well imbalance along $y$}
The measurement of the population imbalance in the $y$-direction as a function of $\varphi_x$ for Fig. 3 and 4 of the main text is performed after an integer or half-integer number of pump cycles, i.e.~$\varphi_x = l \pi$, $l \in \mathbb{Z}$. At these points, the superlattice along $x$ is in the staggered configuration with the maximum energy offset $|\Delta_x| \gg J_x$ and $\delta J_x = 0$. The atoms are thus fully localized on either even or odd sites along $x$ for $\varphi_x = 2 l \pi$ or $\varphi_x = (2l + 1) \pi$, respectively. The four-site unit cell of the 2D superlattice therefore effectively reduces to a double well along $y$. 

For singly-occupied double wells, the expected imbalance in the $y$-direction for atoms in the ground ($\mathcal{I}_y^{\mathrm{gs}}$) and first excited state ($\mathcal{I}_y^{\mathrm{exc}}$) can then be calculated from the single-particle double well Hamiltonian:
\begin{equation}
\label{eq:S23b}
\hat{H}\sub{DW}^{(1)} (\varphi_y)
=
\left( \begin{array}{cc} \Delta_y (\varphi_y)/2 & - J_y^0 (\varphi_y) \\  - J_y^0 (\varphi_y)  & -\Delta_y (\varphi_y)/2 \end{array} \right) 
\end{equation}
with $J_y^0 (\varphi_y) = J_y(\varphi_y) + \delta J_y (\varphi_y)/2$ and using the Fock basis for the atom on the even and odd site, $\ket{1,0}$ and $\ket{0,1}$, respectively. 

Correspondingly, the imbalance for the ground state of a doubly-occupied double well ($\mathcal{I}_y^{\mathrm{2,gs}}$) can be determined using the two-particle double well Hamiltonian:
\begin{equation}
\label{eq:S23c}
\hat{H}\sub{DW}^{(2)} (\varphi_y)
=
\left( \begin{array}{ccc} U + \Delta_y & - \sqrt{2} J_y^0  & 0 \\  -\sqrt{2} J_y^0 & 0 & -\sqrt{2} J_y^0 \\ 0 & -\sqrt{2} J_y^0 & U - \Delta_y  \end{array} \right) 
\end{equation}
in the Fock basis $\{\ket{2,0},\ket{1,1},\ket{0,2} \}$. Here, $U$ denotes the on-site interaction energy for two atoms localized on the same lattice site.

\section{Removal of doubly-occupied sites}
After preparing the Mott insulator with unit filling in the long lattices, sites containing two atoms are converted to singly-occupied ones using microwave-dressed spin-changing collisions \cite{Widera:2005_SI} and a resonant optical push-out pulse. For this, the lattice depths are increased to $V\sub{s,x} = 70(2) E\sub{r,s}$, $V\sub{l,x} = 30(1) E\sub{r,s}$, $V\sub{l,y} = 70(2) E\sub{r,l}$ and $V_{z} = 100(3) E\sub{z}$ in 5\,ms to maximize the on-site interaction energy. The atoms, which are initially in the $(F=1, m_F = -1)$ hyperfine state, are converted to $(F=1, m_F = 0)$ with an adiabatic radio-frequency transfer. By ramping a magnetic offset field in the presence of a microwave field, a Landau-Zener sweep is performed that adiabatically converts pairs of $m_F = 0$ atoms on the same lattice site to an $m_F = +1$ and an $m_F = -1$ atom via coherent spin-changing collisions. Subsequently, the $m_F = -1$ atoms are removed by an adiabatic microwave transfer to $(F=2, m_F = -2)$ followed by a resonant optical pulse after lowering the lattices to $V\sub{s,x} = 0 E\sub{r,s}$, $V\sub{l,x} = 30(1) E\sub{r,l}$, $V\sub{l,y} = 40(1) E\sub{r,l}$ and $V_{z} = 40(1) E\sub{z}$.

\section{Measurement of band excitations}
Band excitations in the $y$-direction are measured by adiabatically ramping the superlattice phase $\varphi_y^{(0)}$ from its initial value to $\pi/2 \pm 0.156(5)\pi$ and subsequently increasing the short lattice depth to $V\sub{s,y} = 40(1) E\sub{r,s}$. In this lattice configuration, ground state atoms on both singly- and doubly-occupied plaquettes are fully localized on the lower-lying site along $y$ due to the large double well tilt $\Delta_y$ and the suppression of tunnelling $J_y^0 \rightarrow 0$. Atoms in the excited band along $y$, on the other hand, localize on the higher-lying site and can be detected directly by measuring the resulting double well imbalance.

\section{Detection of doubly-occupied plaquettes}
The doublon fraction can be determined by taking advantage of the fact that two atoms in the same double well localize on the lower-lying site only at much larger double well tilts than a single atom due to the repulsive on-site interaction. For this, the double wells along $y$ are first merged to a single site by removing the short lattice and increasing the long lattice to $V\sub{l,y} = 30(1) E\sub{r,l}$ within 5\,ms. At the same time, the orthogonal lattice depths are ramped up to $V\sub{s,x} = 70(2) E\sub{r,s}$ and $V_{z} = 100(3) E\sub{r,z}$ to increase the interaction energy. After that, $\varphi_y^{(0)}$ is shifted adiabatically to either $0.474(5)\pi$ or $0.431(5)\pi$ and the sites are split into double wells again by ramping up the short lattice to $V\sub{s,y} = 40(1) E\sub{r,s}$. At $\varphi_y^{(0)} = 0.431\pi$, both single atoms and doublons are fully localized on the lower-lying site. At $\varphi_y^{(0)} = 0.474\pi$, on the other hand, single atoms are still very well localized on the lower site, but two atoms in the same double well localize on different sites due to the large interaction energy of $U/h = 5.4$\,kHz. By determining the site occupations for both phases, one can thus infer the doublon fraction from the difference in the even-odd imbalance between the two measurements.

\section{Alignment of the tilted superlattice}
Each optical lattice is created by retroreflecting a laser beam, which is focussed onto the atoms by a lens on either side of the cloud. For the superlattices, the incoming beams of the short and long lattice are overlapped with a dichroic mirror in front of the first lens. In order to control the tilt angle $\theta$ of the long lattice along $y$, a glass block is placed in the beam path prior to the overlapping. By rotating this glass block, a parallel displacement of the incoming beam can be induced, which is then converted into an angle $\theta$ relative to the short lattice beam at the first lens. The two beams intersect at the focus point of the lens, which corresponds to the position of the atom cloud. After passing through the second lens behind the cloud, both beams are retroreflected by the same mirror. The counterpropagating beams travel along the paths of the incoming beams, thereby creating the lattice potentials with the same relative angle $\theta$.

\section{Determination of the angle $\theta$}
When the long lattice in the $y$-direction is tilted by an angle $\theta$ with respect to the short lattice, the phase of the superlattice along $y$ depends on the position along $x$. This leads to a modification of the on-site potential, which for small angles can be approximated as a linear gradient along the $x$-axis, pointing in opposite directions on even and odd sites in $y$: $\Delta_y^{m_y} (\varphi_y) \approx \Delta_y^{m_y}(\varphi_y^{(0)}) + (-1)^{m_y} \delta \, m_x$. The strength of the gradient is given by $\delta = \pi d\sub{s}/d\sub{l} \, \partial \Delta_y / \partial \varphi_y \big|_{x=0} \, \theta$ for a given superlattice phase $\varphi_y^{(0)}$ and can thus be used to determine $\theta$. In order to do this in the experiment, a superfluid is prepared at $\mathbf{k} = 0$ in a 2D lattice with $V\sub{s,x} = 13.0(4) E\sub{r,s}$, $V\sub{s,y} = 20.0(6) E\sub{r,s}$ and $V\sub{l,y} = 70(2) E\sub{r,l}$. The superlattice phase $\varphi_y^{(0)}$ is set to either $0.344(5)\pi$ or $0.656(5)\pi$ such that the atoms are fully localized on even or odd sites along $y$, respectively. The Bloch oscillations induced by the gradient are probed by measuring the momentum distribution of the atoms after a variable hold time. The angle $\theta$ is then calculated from the average Bloch oscillation period of both phases to minimize the influence of additional residual gradients.

\section{Non-linear response versus lattice depth}

\begin{figure}[t!]
\includegraphics{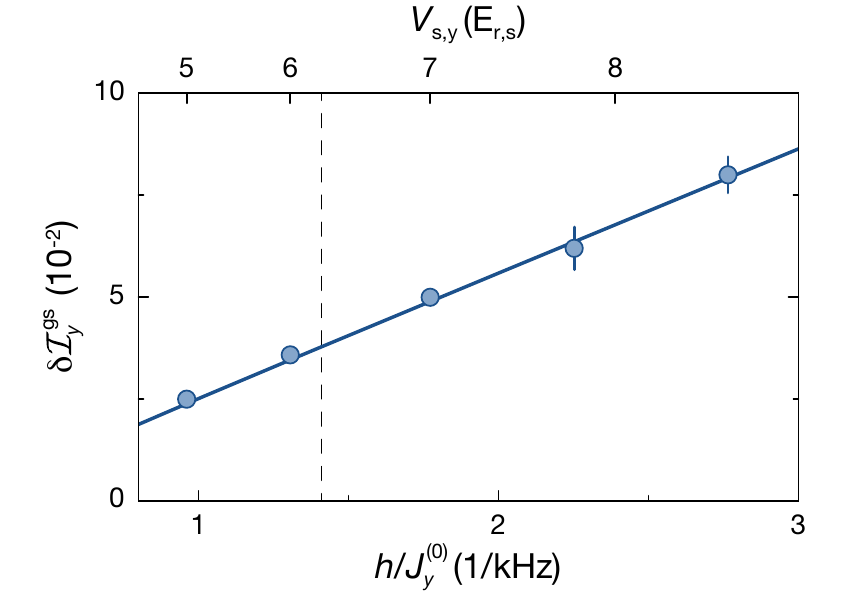}
\caption{Slope of the non-linear response at $\varphi_y^{(0)} = 0.500(5)\pi$ and $\theta = 0.54(3)\, \mathrm{mrad}$ versus short lattice depth along $y$ with all other lattice parameters as in Fig.~3 and 4 of the main text. $J_y^{(0)} = J_y(\varphi_y^{(0)}) + \delta J_y (\varphi_y^{(0)})/2$ with $\varphi_y^{(0)} = \pi/2$ is the maximum intra-double-well tunnelling rate along $y$, which is calculated from the corresponding lattice depth. The solid line indicates the theoretically expected slope and the error bars show the fit error for the slope. The dashed line at $V\sub{s,y} = 6.25 E\sub{r,s}$ marks the point at which a topological transition occurs in the first excited subband along $y$, indicating the transition between the Landau regime for $V\sub{s,y} < 6.25 E\sub{r,s}$ and the Hofstadter regime for $V\sub{s,y} > 6.25 E\sub{r,s}$.
 \label{fig:S3}}
\end{figure}

The technique for detecting the non-linear response with site-resolved band mapping introduced in the main text allows to accurately determine the slope over a wide range of lattice parameters. To demonstrate this, we measure the slope of the non-linear response at $\varphi_y^{(0)} = 0.500(5)\pi$ and $\theta = 0.54(3)\, \mathrm{mrad}$ for various values of the transverse short lattice depth $V\sub{s,y}$ (\extfig{fig:S3}). As expected, the slope increases with larger depths as the band gap decreases and the Berry curvature $\Omega^y$ becomes more and more localized around $\varphi_y^{(0)} = (l + 1/2) \pi$ with $l \in \mathbb{Z}$. 

At $V\sub{s,y} = 6.25 E\sub{r,s}$, the first and second excited subband along $y$ touch for $\varphi_y^{(0)} = l \pi$, leading to a topological transition where the signs of the first and second Chern number of the first excited subband change from $+1$ for $V\sub{s,y} < 6.25 E\sub{r,s}$ to $-1$ for $V\sub{s,y} > 6.25 E\sub{r,s}$. This corresponds to a transition between the Landau and Hofstadter regimes \cite{Lohse:2016_SI}. For the lowest band, the two regimes are topologically equivalent and the atoms thus move in the same direction. In both limits, the experimentally determined slope matches very well with the one expected in an ideal system. This nicely illustrates that the transport properties of the lowest band can be extracted correctly in both regimes, even in the presence of atoms in the first excited band. 

\section{Direct Determination of the second Chern number}
The method for determining the second Chern number from the measurement of the non-linear response versus $\varphi_y^{(0)}$ presented in the main text relies on prior knowledge of the response of an ideal system, both for the ground state and the excited states. While this significantly improves the accuracy for $\nu_2^{\mathrm{exp}}$, it can also be determined directly from the measured double well imbalance $\mathcal{I}_y (\varphi_x)$ without any additional information about the system. To this end, the average change of the imbalance per cycle for the entire cloud, $\delta \mathcal{I}_y (\varphi_y^{(0)})$, is obtained from a linear fit of the differential imbalance $\mathcal{I}_y (\varphi_x) - \mathcal{I}_y (-\varphi_x)$ for each value of $\varphi_y^{(0)}$. The influence of the excitations can be reduced by restricting the fitting region to a small number of pump cycles. The response of an infinite system is reconstructed by averaging $\delta \mathcal{I}_y (\varphi_y^{(0)})$ over $\varphi_y^{(0)}$ using linear interpolation between the data points. When taking into account all points with $\varphi_x/(2\pi) \leq 3$ as well as the finite pumping efficiency along $x$, this gives $\nu_2^{\mathrm{exp}} = 0.94(19)$ for the data from Fig.~3. Note that the linear interpolation for the discrete sampling used in Fig.~3c leads to a systematic shift of $\nu_2^{\mathrm{exp}}$ by +0.05.

\bibliographystyle{bosons}
\input{2D_Pumping_SI_Bibliography.bbl}


\end{document}

%% file: 2D_Pumping_SI_Bibliography.bbl
%

%% file: 2D_Pumping_arXiv.bbl
\begin{thebibliography}{36}%
\makeatletter
\providecommand \@ifxundefined [1]{%
 \@ifx{#1\undefined}
}%
\providecommand \@ifnum [1]{%
 \ifnum #1\expandafter \@firstoftwo
 \else \expandafter \@secondoftwo
 \fi
}%
\providecommand \@ifx [1]{%
 \ifx #1\expandafter \@firstoftwo
 \else \expandafter \@secondoftwo
 \fi
}%
\providecommand \natexlab [1]{#1}%
\providecommand \enquote  [1]{``#1''}%
\providecommand \bibnamefont  [1]{#1}%
\providecommand \bibfnamefont [1]{#1}%
\providecommand \citenamefont [1]{#1}%
\providecommand \href@noop [0]{\@secondoftwo}%
\providecommand \href [0]{\begingroup \@sanitize@url \@href}%
\providecommand \@href[1]{\@@startlink{#1}\@@href}%
\providecommand \@@href[1]{\endgroup#1\@@endlink}%
\providecommand \@sanitize@url [0]{\catcode `\\12\catcode `\$12\catcode
  `\&12\catcode `\#12\catcode `\^12\catcode `\_12\catcode `\%12\relax}%
\providecommand \@@startlink[1]{}%
\providecommand \@@endlink[0]{}%
\providecommand \url  [0]{\begingroup\@sanitize@url \@url }%
\providecommand \@url [1]{\endgroup\@href {#1}{\urlprefix }}%
\providecommand \urlprefix  [0]{URL }%
\providecommand \Eprint [0]{\href }%
\providecommand \doibase [0]{http://dx.doi.org/}%
\providecommand \selectlanguage [0]{\@gobble}%
\providecommand \bibinfo  [0]{\@secondoftwo}%
\providecommand \bibfield  [0]{\@secondoftwo}%
\providecommand \translation [1]{[#1]}%
\providecommand \BibitemOpen [0]{}%
\providecommand \bibitemStop [0]{}%
\providecommand \bibitemNoStop [0]{.\EOS\space}%
\providecommand \EOS [0]{\spacefactor3000\relax}%
\providecommand \BibitemShut  [1]{\csname bibitem#1\endcsname}%
\let\auto@bib@innerbib\@empty
\bibitem [{\citenamefont {Klitzing et~al.}(1980)\citenamefont {Klitzing},
  \citenamefont {Dorda},\ and\ \citenamefont {Pepper}}]{Klitzing:1980}%
  \BibitemOpen
  \bibfield  {author} {\bibinfo {author} {\bibfnamefont {K.~v.}\ \bibnamefont
  {Klitzing}}, \bibinfo {author} {\bibfnamefont {G.}~\bibnamefont {Dorda}},\
  \bibnamefont {and}\ \bibinfo {author} {\bibfnamefont {M.}~\bibnamefont
  {Pepper}},\ }\href {\doibase 10.1103/PhysRevLett.45.494} {\bibfield
  {journal} {\bibinfo  {journal} {Phys. Rev. Lett.}\ }\textbf {\bibinfo
  {volume} {45}},\ \bibinfo {pages} {494}} (\bibinfo {year} {1980})\BibitemShut
  {NoStop}%
\bibitem [{\citenamefont {Thouless et~al.}(1982)\citenamefont {Thouless},
  \citenamefont {Kohmoto}, \citenamefont {Nightingale},\ and\ \citenamefont
  {{den Nijs}}}]{Thouless:1982}%
  \BibitemOpen
  \bibfield  {author} {\bibinfo {author} {\bibfnamefont {D.~J.}\ \bibnamefont
  {Thouless}}, \bibinfo {author} {\bibfnamefont {M.}~\bibnamefont {Kohmoto}},
  \bibinfo {author} {\bibfnamefont {M.~P.}\ \bibnamefont {Nightingale}},\
  \bibnamefont {and}\ \bibinfo {author} {\bibfnamefont {M.}~\bibnamefont {{den
  Nijs}}},\ }\href {\doibase 10.1103/PhysRevLett.49.405} {\bibfield  {journal}
  {\bibinfo  {journal} {Phys. Rev. Lett.}\ }\textbf {\bibinfo {volume} {49}},\
  \bibinfo {pages} {405}} (\bibinfo {year} {1982})\BibitemShut {NoStop}%
\bibitem [{\citenamefont {Yang}(1978)}]{Yang:1978}%
  \BibitemOpen
  \bibfield  {author} {\bibinfo {author} {\bibfnamefont {C.~N.}\ \bibnamefont
  {Yang}},\ }\href {\doibase 10.1063/1.523506} {\bibfield  {journal} {\bibinfo
  {journal} {J. Math. Phys.}\ }\textbf {\bibinfo {volume} {19}},\ \bibinfo
  {pages} {320}} (\bibinfo {year} {1978})\BibitemShut {NoStop}%
\bibitem [{\citenamefont {Zhang\ and\ Hu}(2001)\citenamefont {Zhang}\ and\
  \citenamefont {Hu}}]{Zhang:2001}%
  \BibitemOpen
  \bibfield  {author} {\bibinfo {author} {\bibfnamefont {S.-C.}\ \bibnamefont
  {Zhang}}\ \bibnamefont {and}\ \bibinfo {author} {\bibfnamefont
  {J.}~\bibnamefont {Hu}},\ }\href {\doibase 10.1126/science.294.5543.823}
  {\bibfield  {journal} {\bibinfo  {journal} {Science}\ }\textbf {\bibinfo
  {volume} {294}},\ \bibinfo {pages} {823}} (\bibinfo {year}
  {2001})\BibitemShut {NoStop}%
\bibitem [{\citenamefont {Thouless}(1983)}]{Thouless:1983}%
  \BibitemOpen
  \bibfield  {author} {\bibinfo {author} {\bibfnamefont {D.~J.}\ \bibnamefont
  {Thouless}},\ }\href {\doibase 10.1103/PhysRevB.27.6083} {\bibfield
  {journal} {\bibinfo  {journal} {Phys. Rev. B}\ }\textbf {\bibinfo {volume}
  {27}},\ \bibinfo {pages} {6083}} (\bibinfo {year} {1983})\BibitemShut
  {NoStop}%
\bibitem [{\citenamefont {Kraus et~al.}(2013)\citenamefont {Kraus},
  \citenamefont {Ringel},\ and\ \citenamefont {Zilberberg}}]{Kraus:2013}%
  \BibitemOpen
  \bibfield  {author} {\bibinfo {author} {\bibfnamefont {Y.~E.}\ \bibnamefont
  {Kraus}}, \bibinfo {author} {\bibfnamefont {Z.}~\bibnamefont {Ringel}},\
  \bibnamefont {and}\ \bibinfo {author} {\bibfnamefont {O.}~\bibnamefont
  {Zilberberg}},\ }\href {\doibase 10.1103/PhysRevLett.111.226401} {\bibfield
  {journal} {\bibinfo  {journal} {Phys. Rev. Lett.}\ }\textbf {\bibinfo
  {volume} {111}},\ \bibinfo {pages} {226401}} (\bibinfo {year}
  {2013})\BibitemShut {NoStop}%
\bibitem [{\citenamefont {Yang\ and\ Mills}(1954)\citenamefont {Yang}\ and\
  \citenamefont {Mills}}]{Yang:1954}%
  \BibitemOpen
  \bibfield  {author} {\bibinfo {author} {\bibfnamefont {C.~N.}\ \bibnamefont
  {Yang}}\ \bibnamefont {and}\ \bibinfo {author} {\bibfnamefont {R.~L.}\
  \bibnamefont {Mills}},\ }\href {\doibase 10.1103/PhysRev.96.191} {\bibfield
  {journal} {\bibinfo  {journal} {Phys. Rev.}\ }\textbf {\bibinfo {volume}
  {96}},\ \bibinfo {pages} {191}} (\bibinfo {year} {1954})\BibitemShut
  {NoStop}%
\bibitem [{\citenamefont {Qi\ and\ Zhang}(2011)\citenamefont {Qi}\ and\
  \citenamefont {Zhang}}]{Qi:2011}%
  \BibitemOpen
  \bibfield  {author} {\bibinfo {author} {\bibfnamefont {X.-L.}\ \bibnamefont
  {Qi}}\ \bibnamefont {and}\ \bibinfo {author} {\bibfnamefont {S.-C.}\
  \bibnamefont {Zhang}},\ }\href {\doibase 10.1103/RevModPhys.83.1057}
  {\bibfield  {journal} {\bibinfo  {journal} {Rev. Mod. Phys.}\ }\textbf
  {\bibinfo {volume} {83}},\ \bibinfo {pages} {1057}} (\bibinfo {year}
  {2011})\BibitemShut {NoStop}%
\bibitem [{\citenamefont {Nayak et~al.}(2008)\citenamefont {Nayak},
  \citenamefont {Simon}, \citenamefont {Stern}, \citenamefont {Freedman},\ and\
  \citenamefont {Das~Sarma}}]{Nayak:2008}%
  \BibitemOpen
  \bibfield  {author} {\bibinfo {author} {\bibfnamefont {C.}~\bibnamefont
  {Nayak}}, \bibinfo {author} {\bibfnamefont {S.~H.}\ \bibnamefont {Simon}},
  \bibinfo {author} {\bibfnamefont {A.}~\bibnamefont {Stern}}, \bibinfo
  {author} {\bibfnamefont {M.}~\bibnamefont {Freedman}},\ \bibnamefont {and}\
  \bibinfo {author} {\bibfnamefont {S.}~\bibnamefont {Das~Sarma}},\ }\href
  {\doibase 10.1103/RevModPhys.80.1083} {\bibfield  {journal} {\bibinfo
  {journal} {Rev. Mod. Phys.}\ }\textbf {\bibinfo {volume} {80}},\ \bibinfo
  {pages} {1083}} (\bibinfo {year} {2008})\BibitemShut {NoStop}%
\bibitem [{\citenamefont {Laughlin}(1981)}]{Laughlin:1981}%
  \BibitemOpen
  \bibfield  {author} {\bibinfo {author} {\bibfnamefont {R.~B.}\ \bibnamefont
  {Laughlin}},\ }\href {\doibase 10.1103/PhysRevB.23.5632} {\bibfield
  {journal} {\bibinfo  {journal} {Phys. Rev. B}\ }\textbf {\bibinfo {volume}
  {23}},\ \bibinfo {pages} {5632}} (\bibinfo {year} {1981})\BibitemShut
  {NoStop}%
\bibitem [{\citenamefont {Lu et~al.}(2015)\citenamefont {Lu}, \citenamefont
  {Wang}, \citenamefont {Ye}, \citenamefont {Ran}, \citenamefont {Fu},
  \citenamefont {Joannopoulos},\ and\ \citenamefont {Solja{\v
  c}i{\'c}}}]{Lu:2015}%
  \BibitemOpen
  \bibfield  {author} {\bibinfo {author} {\bibfnamefont {L.}~\bibnamefont
  {Lu}}, \bibinfo {author} {\bibfnamefont {Z.}~\bibnamefont {Wang}}, \bibinfo
  {author} {\bibfnamefont {D.}~\bibnamefont {Ye}}, \bibnamefont {et~al.},\
  }\href {\doibase 10.1126/science.aaa9273} {\bibfield  {journal} {\bibinfo
  {journal} {Science}\ }\textbf {\bibinfo {volume} {349}},\ \bibinfo {pages}
  {622}} (\bibinfo {year} {2015})\BibitemShut {NoStop}%
\bibitem [{\citenamefont {Xu et~al.}(2015)\citenamefont {Xu}, \citenamefont
  {Belopolski}, \citenamefont {Alidoust}, \citenamefont {Neupane},
  \citenamefont {Bian}, \citenamefont {Zhang}, \citenamefont {Sankar},
  \citenamefont {Chang}, \citenamefont {Yuan}, \citenamefont {Lee},
  \citenamefont {Huang}, \citenamefont {Zheng}, \citenamefont {Ma},
  \citenamefont {Sanchez}, \citenamefont {Wang}, \citenamefont {Bansil},
  \citenamefont {Chou}, \citenamefont {Shibayev}, \citenamefont {Lin},
  \citenamefont {Jia},\ and\ \citenamefont {Hasan}}]{Xu:2015}%
  \BibitemOpen
  \bibfield  {author} {\bibinfo {author} {\bibfnamefont {S.-Y.}\ \bibnamefont
  {Xu}}, \bibinfo {author} {\bibfnamefont {I.}~\bibnamefont {Belopolski}},
  \bibinfo {author} {\bibfnamefont {N.}~\bibnamefont {Alidoust}}, \bibnamefont
  {et~al.},\ }\href {\doibase 10.1126/science.aaa9297} {\bibfield  {journal}
  {\bibinfo  {journal} {Science}\ }\textbf {\bibinfo {volume} {349}},\ \bibinfo
  {pages} {613}} (\bibinfo {year} {2015})\BibitemShut {NoStop}%
\bibitem [{\citenamefont {Hsieh et~al.}(2008)\citenamefont {Hsieh},
  \citenamefont {Qian}, \citenamefont {Wray}, \citenamefont {Xia},
  \citenamefont {Hor}, \citenamefont {Cava},\ and\ \citenamefont
  {Hasan}}]{Hsieh:2008}%
  \BibitemOpen
  \bibfield  {author} {\bibinfo {author} {\bibfnamefont {D.}~\bibnamefont
  {Hsieh}}, \bibinfo {author} {\bibfnamefont {D.}~\bibnamefont {Qian}},
  \bibinfo {author} {\bibfnamefont {L.}~\bibnamefont {Wray}}, \bibnamefont
  {et~al.},\ }\href {\doibase 10.1038/nature06843} {\bibfield  {journal}
  {\bibinfo  {journal} {Nature}\ }\textbf {\bibinfo {volume} {452}},\ \bibinfo
  {pages} {970}} (\bibinfo {year} {2008})\BibitemShut {NoStop}%
\bibitem [{\citenamefont {Qi et~al.}(2008)\citenamefont {Qi}, \citenamefont
  {Hughes},\ and\ \citenamefont {Zhang}}]{Qi:2008}%
  \BibitemOpen
  \bibfield  {author} {\bibinfo {author} {\bibfnamefont {X.-L.}\ \bibnamefont
  {Qi}}, \bibinfo {author} {\bibfnamefont {T.~L.}\ \bibnamefont {Hughes}},\
  \bibnamefont {and}\ \bibinfo {author} {\bibfnamefont {S.-C.}\ \bibnamefont
  {Zhang}},\ }\href {\doibase 10.1103/PhysRevB.78.195424} {\bibfield  {journal}
  {\bibinfo  {journal} {Phys. Rev. B}\ }\textbf {\bibinfo {volume} {78}},\
  \bibinfo {pages} {195424}} (\bibinfo {year} {2008})\BibitemShut {NoStop}%
\bibitem [{\citenamefont {Edge et~al.}(2012)\citenamefont {Edge}, \citenamefont
  {Tworzyd\l{}o},\ and\ \citenamefont {Beenakker}}]{Edge:2012}%
  \BibitemOpen
  \bibfield  {author} {\bibinfo {author} {\bibfnamefont {J.~M.}\ \bibnamefont
  {Edge}}, \bibinfo {author} {\bibfnamefont {J.}~\bibnamefont {Tworzyd\l{}o}},\
  \bibnamefont {and}\ \bibinfo {author} {\bibfnamefont {C.~W.~J.}\ \bibnamefont
  {Beenakker}},\ }\href {\doibase 10.1103/PhysRevLett.109.135701} {\bibfield
  {journal} {\bibinfo  {journal} {Phys. Rev. Lett.}\ }\textbf {\bibinfo
  {volume} {109}},\ \bibinfo {pages} {135701}} (\bibinfo {year}
  {2012})\BibitemShut {NoStop}%
\bibitem [{\citenamefont {Li et~al.}(2013)\citenamefont {Li}, \citenamefont
  {Zhang},\ and\ \citenamefont {Wu}}]{Li:2013}%
  \BibitemOpen
  \bibfield  {author} {\bibinfo {author} {\bibfnamefont {Y.}~\bibnamefont
  {Li}}, \bibinfo {author} {\bibfnamefont {S.-C.}\ \bibnamefont {Zhang}},\
  \bibnamefont {and}\ \bibinfo {author} {\bibfnamefont {C.}~\bibnamefont
  {Wu}},\ }\href {\doibase 10.1103/PhysRevLett.111.186803} {\bibfield
  {journal} {\bibinfo  {journal} {Phys. Rev. Lett.}\ }\textbf {\bibinfo
  {volume} {111}},\ \bibinfo {pages} {186803}} (\bibinfo {year}
  {2013})\BibitemShut {NoStop}%
\bibitem [{\citenamefont {Goldman et~al.}(2016)\citenamefont {Goldman},
  \citenamefont {Budich},\ and\ \citenamefont {Zoller}}]{Goldman:2016}%
  \BibitemOpen
  \bibfield  {author} {\bibinfo {author} {\bibfnamefont {N.}~\bibnamefont
  {Goldman}}, \bibinfo {author} {\bibfnamefont {J.~C.}\ \bibnamefont
  {Budich}},\ \bibnamefont {and}\ \bibinfo {author} {\bibfnamefont
  {P.}~\bibnamefont {Zoller}},\ }\href {http://dx.doi.org/10.1038/nphys3803}
  {\bibfield  {journal} {\bibinfo  {journal} {Nat. Phys.}\ }\textbf {\bibinfo
  {volume} {12}},\ \bibinfo {pages} {639}} (\bibinfo {year} {2016})\BibitemShut
  {NoStop}%
\bibitem [{\citenamefont {Lu et~al.}(2014)\citenamefont {Lu}, \citenamefont
  {Joannopoulos},\ and\ \citenamefont {Soljacic}}]{Lu:2014}%
  \BibitemOpen
  \bibfield  {author} {\bibinfo {author} {\bibfnamefont {L.}~\bibnamefont
  {Lu}}, \bibinfo {author} {\bibfnamefont {J.~D.}\ \bibnamefont
  {Joannopoulos}},\ \bibnamefont {and}\ \bibinfo {author} {\bibfnamefont
  {M.}~\bibnamefont {Soljacic}},\ }\href
  {http://dx.doi.org/10.1038/nphoton.2014.248} {\bibfield  {journal} {\bibinfo
  {journal} {Nat. Photon.}\ }\textbf {\bibinfo {volume} {8}},\ \bibinfo {pages}
  {821}} (\bibinfo {year} {2014})\BibitemShut {NoStop}%
\bibitem [{\citenamefont {Mancini et~al.}(2015)\citenamefont {Mancini},
  \citenamefont {Pagano}, \citenamefont {Cappellini}, \citenamefont {Livi},
  \citenamefont {Rider}, \citenamefont {Catani}, \citenamefont {Sias},
  \citenamefont {Zoller}, \citenamefont {Inguscio}, \citenamefont {Dalmonte},\
  and\ \citenamefont {Fallani}}]{Mancini:2015}%
  \BibitemOpen
  \bibfield  {author} {\bibinfo {author} {\bibfnamefont {M.}~\bibnamefont
  {Mancini}}, \bibinfo {author} {\bibfnamefont {G.}~\bibnamefont {Pagano}},
  \bibinfo {author} {\bibfnamefont {G.}~\bibnamefont {Cappellini}},
  \bibnamefont {et~al.},\ }\href {\doibase 10.1126/science.aaa8736} {\bibfield
  {journal} {\bibinfo  {journal} {Science}\ }\textbf {\bibinfo {volume}
  {349}},\ \bibinfo {pages} {1510}} (\bibinfo {year} {2015})\BibitemShut
  {NoStop}%
\bibitem [{\citenamefont {Stuhl et~al.}(2015)\citenamefont {Stuhl},
  \citenamefont {Lu}, \citenamefont {Aycock}, \citenamefont {Genkina},\ and\
  \citenamefont {Spielman}}]{Stuhl:2015}%
  \BibitemOpen
  \bibfield  {author} {\bibinfo {author} {\bibfnamefont {B.~K.}\ \bibnamefont
  {Stuhl}}, \bibinfo {author} {\bibfnamefont {H.-I.}\ \bibnamefont {Lu}},
  \bibinfo {author} {\bibfnamefont {L.~M.}\ \bibnamefont {Aycock}}, \bibinfo
  {author} {\bibfnamefont {D.}~\bibnamefont {Genkina}},\ \bibnamefont {and}\
  \bibinfo {author} {\bibfnamefont {I.~B.}\ \bibnamefont {Spielman}},\ }\href
  {\doibase 10.1126/science.aaa8515} {\bibfield  {journal} {\bibinfo  {journal}
  {Science}\ }\textbf {\bibinfo {volume} {349}},\ \bibinfo {pages} {1514}}
  (\bibinfo {year} {2015})\BibitemShut {NoStop}%
\bibitem [{\citenamefont {Price et~al.}(2015)\citenamefont {Price},
  \citenamefont {Zilberberg}, \citenamefont {Ozawa}, \citenamefont
  {Carusotto},\ and\ \citenamefont {Goldman}}]{Price:2015}%
  \BibitemOpen
  \bibfield  {author} {\bibinfo {author} {\bibfnamefont {H.~M.}\ \bibnamefont
  {Price}}, \bibinfo {author} {\bibfnamefont {O.}~\bibnamefont {Zilberberg}},
  \bibinfo {author} {\bibfnamefont {T.}~\bibnamefont {Ozawa}}, \bibinfo
  {author} {\bibfnamefont {I.}~\bibnamefont {Carusotto}},\ \bibnamefont {and}\
  \bibinfo {author} {\bibfnamefont {N.}~\bibnamefont {Goldman}},\ }\href
  {\doibase 10.1103/PhysRevLett.115.195303} {\bibfield  {journal} {\bibinfo
  {journal} {Phys. Rev. Lett.}\ }\textbf {\bibinfo {volume} {115}},\ \bibinfo
  {pages} {195303}} (\bibinfo {year} {2015})\BibitemShut {NoStop}%
\bibitem [{\citenamefont {Ozawa et~al.}(2016)\citenamefont {Ozawa},
  \citenamefont {Price}, \citenamefont {Goldman}, \citenamefont {Zilberberg},\
  and\ \citenamefont {Carusotto}}]{Ozawa:2016}%
  \BibitemOpen
  \bibfield  {author} {\bibinfo {author} {\bibfnamefont {T.}~\bibnamefont
  {Ozawa}}, \bibinfo {author} {\bibfnamefont {H.~M.}\ \bibnamefont {Price}},
  \bibinfo {author} {\bibfnamefont {N.}~\bibnamefont {Goldman}}, \bibinfo
  {author} {\bibfnamefont {O.}~\bibnamefont {Zilberberg}},\ \bibnamefont {and}\
  \bibinfo {author} {\bibfnamefont {I.}~\bibnamefont {Carusotto}},\ }\href
  {\doibase 10.1103/PhysRevA.93.043827} {\bibfield  {journal} {\bibinfo
  {journal} {Phys. Rev. A}\ }\textbf {\bibinfo {volume} {93}},\ \bibinfo
  {pages} {043827}} (\bibinfo {year} {2016})\BibitemShut {NoStop}%
\bibitem [{\citenamefont {Price et~al.}(2016)\citenamefont {Price},
  \citenamefont {Zilberberg}, \citenamefont {Ozawa}, \citenamefont
  {Carusotto},\ and\ \citenamefont {Goldman}}]{Price:2016}%
  \BibitemOpen
  \bibfield  {author} {\bibinfo {author} {\bibfnamefont {H.~M.}\ \bibnamefont
  {Price}}, \bibinfo {author} {\bibfnamefont {O.}~\bibnamefont {Zilberberg}},
  \bibinfo {author} {\bibfnamefont {T.}~\bibnamefont {Ozawa}}, \bibinfo
  {author} {\bibfnamefont {I.}~\bibnamefont {Carusotto}},\ \bibnamefont {and}\
  \bibinfo {author} {\bibfnamefont {N.}~\bibnamefont {Goldman}},\ }\href
  {\doibase 10.1103/PhysRevB.93.245113} {\bibfield  {journal} {\bibinfo
  {journal} {Phys. Rev. B}\ }\textbf {\bibinfo {volume} {93}},\ \bibinfo
  {pages} {245113}} (\bibinfo {year} {2016})\BibitemShut {NoStop}%
\bibitem [{\citenamefont {Sugawa et~al.}(2016)\citenamefont {Sugawa},
  \citenamefont {Salces-Carcoba}, \citenamefont {Perry}, \citenamefont {Yue},\
  and\ \citenamefont {Spielman}}]{Sugawa:2016}%
  \BibitemOpen
  \bibfield  {author} {\bibinfo {author} {\bibfnamefont {S.}~\bibnamefont
  {Sugawa}}, \bibinfo {author} {\bibfnamefont {F.}~\bibnamefont
  {Salces-Carcoba}}, \bibinfo {author} {\bibfnamefont {A.~R.}\ \bibnamefont
  {Perry}}, \bibinfo {author} {\bibfnamefont {Y.}~\bibnamefont {Yue}},\
  \bibnamefont {and}\ \bibinfo {author} {\bibfnamefont {I.~B.}\ \bibnamefont
  {Spielman}},\ }\href {http://arxiv.org/abs/1610.06228} {\Eprint
  {http://arxiv.org/abs/1610.06228} {arXiv:1610.06228 [cond-mat.quant-gas]} }
  (\bibinfo {year} {2016})\BibitemShut {NoStop}%
\bibitem [{\citenamefont {Kraus\ and\ Zilberberg}(2012)\citenamefont {Kraus}\
  and\ \citenamefont {Zilberberg}}]{Kraus:2012b}%
  \BibitemOpen
  \bibfield  {author} {\bibinfo {author} {\bibfnamefont {Y.~E.}\ \bibnamefont
  {Kraus}}\ \bibnamefont {and}\ \bibinfo {author} {\bibfnamefont
  {O.}~\bibnamefont {Zilberberg}},\ }\href {\doibase
  10.1103/PhysRevLett.109.116404} {\bibfield  {journal} {\bibinfo  {journal}
  {Phys. Rev. Lett.}\ }\textbf {\bibinfo {volume} {109}},\ \bibinfo {pages}
  {116404}} (\bibinfo {year} {2012})\BibitemShut {NoStop}%
\bibitem [{\citenamefont {Pothier et~al.}(1992)\citenamefont {Pothier},
  \citenamefont {Lafarge}, \citenamefont {Urbina}, \citenamefont {Esteve},\
  and\ \citenamefont {Devoret}}]{Pothier:1992}%
  \BibitemOpen
  \bibfield  {author} {\bibinfo {author} {\bibfnamefont {H.}~\bibnamefont
  {Pothier}}, \bibinfo {author} {\bibfnamefont {P.}~\bibnamefont {Lafarge}},
  \bibinfo {author} {\bibfnamefont {C.}~\bibnamefont {Urbina}}, \bibinfo
  {author} {\bibfnamefont {D.}~\bibnamefont {Esteve}},\ \bibnamefont {and}\
  \bibinfo {author} {\bibfnamefont {M.~H.}\ \bibnamefont {Devoret}},\ }\href
  {http://stacks.iop.org/0295-5075/17/i=3/a=011} {\bibfield  {journal}
  {\bibinfo  {journal} {EPL}\ }\textbf {\bibinfo {volume} {17}},\ \bibinfo
  {pages} {249}} (\bibinfo {year} {1992})\BibitemShut {NoStop}%
\bibitem [{\citenamefont {Switkes et~al.}(1999)\citenamefont {Switkes},
  \citenamefont {Marcus}, \citenamefont {Campman},\ and\ \citenamefont
  {Gossard}}]{Switkes:1999}%
  \BibitemOpen
  \bibfield  {author} {\bibinfo {author} {\bibfnamefont {M.}~\bibnamefont
  {Switkes}}, \bibinfo {author} {\bibfnamefont {C.~M.}\ \bibnamefont {Marcus}},
  \bibinfo {author} {\bibfnamefont {K.}~\bibnamefont {Campman}},\ \bibnamefont
  {and}\ \bibinfo {author} {\bibfnamefont {A.~C.}\ \bibnamefont {Gossard}},\
  }\href {\doibase 10.1126/science.283.5409.1905} {\bibfield  {journal}
  {\bibinfo  {journal} {Science}\ }\textbf {\bibinfo {volume} {283}},\ \bibinfo
  {pages} {1905}} (\bibinfo {year} {1999})\BibitemShut {NoStop}%
\bibitem [{\citenamefont {Kraus et~al.}(2012)\citenamefont {Kraus},
  \citenamefont {Lahini}, \citenamefont {Ringel}, \citenamefont {Verbin},\ and\
  \citenamefont {Zilberberg}}]{Kraus:2012a}%
  \BibitemOpen
  \bibfield  {author} {\bibinfo {author} {\bibfnamefont {Y.~E.}\ \bibnamefont
  {Kraus}}, \bibinfo {author} {\bibfnamefont {Y.}~\bibnamefont {Lahini}},
  \bibinfo {author} {\bibfnamefont {Z.}~\bibnamefont {Ringel}}, \bibinfo
  {author} {\bibfnamefont {M.}~\bibnamefont {Verbin}},\ \bibnamefont {and}\
  \bibinfo {author} {\bibfnamefont {O.}~\bibnamefont {Zilberberg}},\ }\href
  {\doibase 10.1103/PhysRevLett.109.106402} {\bibfield  {journal} {\bibinfo
  {journal} {Phys. Rev. Lett.}\ }\textbf {\bibinfo {volume} {109}},\ \bibinfo
  {pages} {106402}} (\bibinfo {year} {2012})\BibitemShut {NoStop}%
\bibitem [{\citenamefont {Verbin et~al.}(2015)\citenamefont {Verbin},
  \citenamefont {Zilberberg}, \citenamefont {Lahini}, \citenamefont {Kraus},\
  and\ \citenamefont {Silberberg}}]{Verbin:2015}%
  \BibitemOpen
  \bibfield  {author} {\bibinfo {author} {\bibfnamefont {M.}~\bibnamefont
  {Verbin}}, \bibinfo {author} {\bibfnamefont {O.}~\bibnamefont {Zilberberg}},
  \bibinfo {author} {\bibfnamefont {Y.}~\bibnamefont {Lahini}}, \bibinfo
  {author} {\bibfnamefont {Y.~E.}\ \bibnamefont {Kraus}},\ \bibnamefont {and}\
  \bibinfo {author} {\bibfnamefont {Y.}~\bibnamefont {Silberberg}},\ }\href
  {\doibase 10.1103/PhysRevB.91.064201} {\bibfield  {journal} {\bibinfo
  {journal} {Phys. Rev. B}\ }\textbf {\bibinfo {volume} {91}},\ \bibinfo
  {pages} {064201}} (\bibinfo {year} {2015})\BibitemShut {NoStop}%
\bibitem [{\citenamefont {Lohse et~al.}(2016)\citenamefont {Lohse},
  \citenamefont {Schweizer}, \citenamefont {Zilberberg}, \citenamefont
  {Aidelsburger},\ and\ \citenamefont {Bloch}}]{Lohse:2016}%
  \BibitemOpen
  \bibfield  {author} {\bibinfo {author} {\bibfnamefont {M.}~\bibnamefont
  {Lohse}}, \bibinfo {author} {\bibfnamefont {C.}~\bibnamefont {Schweizer}},
  \bibinfo {author} {\bibfnamefont {O.}~\bibnamefont {Zilberberg}}, \bibinfo
  {author} {\bibfnamefont {M.}~\bibnamefont {Aidelsburger}},\ \bibnamefont
  {and}\ \bibinfo {author} {\bibfnamefont {I.}~\bibnamefont {Bloch}},\ }\href
  {http://dx.doi.org/10.1038/nphys3584} {\bibfield  {journal} {\bibinfo
  {journal} {Nat. Phys.}\ }\textbf {\bibinfo {volume} {12}},\ \bibinfo {pages}
  {350}} (\bibinfo {year} {2016})\BibitemShut {NoStop}%
\bibitem [{\citenamefont {Nakajima et~al.}(2016)\citenamefont {Nakajima},
  \citenamefont {Tomita}, \citenamefont {Taie}, \citenamefont {Ichinose},
  \citenamefont {Ozawa}, \citenamefont {Wang}, \citenamefont {Troyer},\ and\
  \citenamefont {Takahashi}}]{Nakajima:2016}%
  \BibitemOpen
  \bibfield  {author} {\bibinfo {author} {\bibfnamefont {S.}~\bibnamefont
  {Nakajima}}, \bibinfo {author} {\bibfnamefont {T.}~\bibnamefont {Tomita}},
  \bibinfo {author} {\bibfnamefont {S.}~\bibnamefont {Taie}}, \bibnamefont
  {et~al.},\ }\href {http://dx.doi.org/10.1038/nphys3622} {\bibfield  {journal}
  {\bibinfo  {journal} {Nat. Phys.}\ }\textbf {\bibinfo {volume} {12}},\
  \bibinfo {pages} {296}} (\bibinfo {year} {2016})\BibitemShut {NoStop}%
\bibitem [{\citenamefont {Lu et~al.}(2016)\citenamefont {Lu}, \citenamefont
  {Schemmer}, \citenamefont {Aycock}, \citenamefont {Genkina}, \citenamefont
  {Sugawa},\ and\ \citenamefont {Spielman}}]{Lu:2016}%
  \BibitemOpen
  \bibfield  {author} {\bibinfo {author} {\bibfnamefont {H.-I.}\ \bibnamefont
  {Lu}}, \bibinfo {author} {\bibfnamefont {M.}~\bibnamefont {Schemmer}},
  \bibinfo {author} {\bibfnamefont {L.~M.}\ \bibnamefont {Aycock}},
  \bibnamefont {et~al.},\ }\href {\doibase 10.1103/PhysRevLett.116.200402}
  {\bibfield  {journal} {\bibinfo  {journal} {Phys. Rev. Lett.}\ }\textbf
  {\bibinfo {volume} {116}},\ \bibinfo {pages} {200402}} (\bibinfo {year}
  {2016})\BibitemShut {NoStop}%
\bibitem [{\citenamefont {Schweizer et~al.}(2016)\citenamefont {Schweizer},
  \citenamefont {Lohse}, \citenamefont {Citro},\ and\ \citenamefont
  {Bloch}}]{Schweizer:2016}%
  \BibitemOpen
  \bibfield  {author} {\bibinfo {author} {\bibfnamefont {C.}~\bibnamefont
  {Schweizer}}, \bibinfo {author} {\bibfnamefont {M.}~\bibnamefont {Lohse}},
  \bibinfo {author} {\bibfnamefont {R.}~\bibnamefont {Citro}},\ \bibnamefont
  {and}\ \bibinfo {author} {\bibfnamefont {I.}~\bibnamefont {Bloch}},\ }\href
  {\doibase 10.1103/PhysRevLett.117.170405} {\bibfield  {journal} {\bibinfo
  {journal} {Phys. Rev. Lett.}\ }\textbf {\bibinfo {volume} {117}},\ \bibinfo
  {pages} {170405}} (\bibinfo {year} {2016})\BibitemShut {NoStop}%
\bibitem [{\citenamefont {Rice\ and\ Mele}(1982)\citenamefont {Rice}\ and\
  \citenamefont {Mele}}]{Rice:1982}%
  \BibitemOpen
  \bibfield  {author} {\bibinfo {author} {\bibfnamefont {M.~J.}\ \bibnamefont
  {Rice}}\ \bibnamefont {and}\ \bibinfo {author} {\bibfnamefont {E.~J.}\
  \bibnamefont {Mele}},\ }\href {\doibase 10.1103/PhysRevLett.49.1455}
  {\bibfield  {journal} {\bibinfo  {journal} {Phys. Rev. Lett.}\ }\textbf
  {\bibinfo {volume} {49}},\ \bibinfo {pages} {1455}} (\bibinfo {year}
  {1982})\BibitemShut {NoStop}%
\bibitem [{\citenamefont {Marra et~al.}(2015)\citenamefont {Marra},
  \citenamefont {Citro},\ and\ \citenamefont {Ortix}}]{Marra:2015}%
  \BibitemOpen
  \bibfield  {author} {\bibinfo {author} {\bibfnamefont {P.}~\bibnamefont
  {Marra}}, \bibinfo {author} {\bibfnamefont {R.}~\bibnamefont {Citro}},\
  \bibnamefont {and}\ \bibinfo {author} {\bibfnamefont {C.}~\bibnamefont
  {Ortix}},\ }\href {\doibase 10.1103/PhysRevB.91.125411} {\bibfield  {journal}
  {\bibinfo  {journal} {Phys. Rev. B}\ }\textbf {\bibinfo {volume} {91}},\
  \bibinfo {pages} {125411}} (\bibinfo {year} {2015})\BibitemShut {NoStop}%
\bibitem [{\citenamefont {Zhu et~al.}(2013)\citenamefont {Zhu}, \citenamefont
  {Wang}, \citenamefont {Chan},\ and\ \citenamefont {Duan}}]{Zhu:2013}%
  \BibitemOpen
  \bibfield  {author} {\bibinfo {author} {\bibfnamefont {S.-L.}\ \bibnamefont
  {Zhu}}, \bibinfo {author} {\bibfnamefont {Z.-D.}\ \bibnamefont {Wang}},
  \bibinfo {author} {\bibfnamefont {Y.-H.}\ \bibnamefont {Chan}},\ \bibnamefont
  {and}\ \bibinfo {author} {\bibfnamefont {L.-M.}\ \bibnamefont {Duan}},\
  }\href {\doibase 10.1103/PhysRevLett.110.075303} {\bibfield  {journal}
  {\bibinfo  {journal} {Phys. Rev. Lett.}\ }\textbf {\bibinfo {volume} {110}},\
  \bibinfo {pages} {075303}} (\bibinfo {year} {2013})\BibitemShut {NoStop}%
\end{thebibliography}

\begin{thebibliography}{11}%
\makeatletter
\providecommand \@ifxundefined [1]{%
 \@ifx{#1\undefined}
}%
\providecommand \@ifnum [1]{%
 \ifnum #1\expandafter \@firstoftwo
 \else \expandafter \@secondoftwo
 \fi
}%
\providecommand \@ifx [1]{%
 \ifx #1\expandafter \@firstoftwo
 \else \expandafter \@secondoftwo
 \fi
}%
\providecommand \natexlab [1]{#1}%
\providecommand \enquote  [1]{``#1''}%
\providecommand \bibnamefont  [1]{#1}%
\providecommand \bibfnamefont [1]{#1}%
\providecommand \citenamefont [1]{#1}%
\providecommand \href@noop [0]{\@secondoftwo}%
\providecommand \href [0]{\begingroup \@sanitize@url \@href}%
\providecommand \@href[1]{\@@startlink{#1}\@@href}%
\providecommand \@@href[1]{\endgroup#1\@@endlink}%
\providecommand \@sanitize@url [0]{\catcode `\\12\catcode `\$12\catcode
  `\&12\catcode `\#12\catcode `\^12\catcode `\_12\catcode `\%12\relax}%
\providecommand \@@startlink[1]{}%
\providecommand \@@endlink[0]{}%
\providecommand \url  [0]{\begingroup\@sanitize@url \@url }%
\providecommand \@url [1]{\endgroup\@href {#1}{\urlprefix }}%
\providecommand \urlprefix  [0]{URL }%
\providecommand \Eprint [0]{\href }%
\providecommand \doibase [0]{http://dx.doi.org/}%
\providecommand \selectlanguage [0]{\@gobble}%
\providecommand \bibinfo  [0]{\@secondoftwo}%
\providecommand \bibfield  [0]{\@secondoftwo}%
\providecommand \translation [1]{[#1]}%
\providecommand \BibitemOpen [0]{}%
\providecommand \bibitemStop [0]{}%
\providecommand \bibitemNoStop [0]{.\EOS\space}%
\providecommand \EOS [0]{\spacefactor3000\relax}%
\providecommand \BibitemShut  [1]{\csname bibitem#1\endcsname}%
\let\auto@bib@innerbib\@empty
\bibitem [{\citenamefont {Xiao et~al.}(2010)\citenamefont {Xiao}, \citenamefont
  {Chang},\ and\ \citenamefont {Niu}}]{Xiao:2010_SI}%
  \BibitemOpen
  \bibfield  {author} {\bibinfo {author} {\bibfnamefont {D.}~\bibnamefont
  {Xiao}}, \bibinfo {author} {\bibfnamefont {M.-C.}\ \bibnamefont {Chang}},\
  \bibnamefont {and}\ \bibinfo {author} {\bibfnamefont {Q.}~\bibnamefont
  {Niu}},\ }\href {\doibase 10.1103/RevModPhys.82.1959} {\bibfield  {journal}
  {\bibinfo  {journal} {Rev. Mod. Phys.}\ }\textbf {\bibinfo {volume} {82}},\
  \bibinfo {pages} {1959}} (\bibinfo {year} {2010})\BibitemShut {NoStop}%
\bibitem [{\citenamefont {Price et~al.}(2015)\citenamefont {Price},
  \citenamefont {Zilberberg}, \citenamefont {Ozawa}, \citenamefont
  {Carusotto},\ and\ \citenamefont {Goldman}}]{Price:2015_SI}%
  \BibitemOpen
  \bibfield  {author} {\bibinfo {author} {\bibfnamefont {H.~M.}\ \bibnamefont
  {Price}}, \bibinfo {author} {\bibfnamefont {O.}~\bibnamefont {Zilberberg}},
  \bibinfo {author} {\bibfnamefont {T.}~\bibnamefont {Ozawa}}, \bibinfo
  {author} {\bibfnamefont {I.}~\bibnamefont {Carusotto}},\ \bibnamefont {and}\
  \bibinfo {author} {\bibfnamefont {N.}~\bibnamefont {Goldman}},\ }\href
  {\doibase 10.1103/PhysRevLett.115.195303} {\bibfield  {journal} {\bibinfo
  {journal} {Phys. Rev. Lett.}\ }\textbf {\bibinfo {volume} {115}},\ \bibinfo
  {pages} {195303}} (\bibinfo {year} {2015})\BibitemShut {NoStop}%
\bibitem [{\citenamefont {Kraus\ and\ Zilberberg}(2012)\citenamefont {Kraus}\
  and\ \citenamefont {Zilberberg}}]{Kraus:2012b_SI}%
  \BibitemOpen
  \bibfield  {author} {\bibinfo {author} {\bibfnamefont {Y.~E.}\ \bibnamefont
  {Kraus}}\ \bibnamefont {and}\ \bibinfo {author} {\bibfnamefont
  {O.}~\bibnamefont {Zilberberg}},\ }\href {\doibase
  10.1103/PhysRevLett.109.116404} {\bibfield  {journal} {\bibinfo  {journal}
  {Phys. Rev. Lett.}\ }\textbf {\bibinfo {volume} {109}},\ \bibinfo {pages}
  {116404}} (\bibinfo {year} {2012})\BibitemShut {NoStop}%
\bibitem [{\citenamefont {Harper}(1955)}]{Harper:1955_SI}%
  \BibitemOpen
  \bibfield  {author} {\bibinfo {author} {\bibfnamefont {P.~G.}\ \bibnamefont
  {Harper}},\ }\href {http://stacks.iop.org/0370-1298/68/i=10/a=305} {\bibfield
   {journal} {\bibinfo  {journal} {Proc. Phys. Soc. A}\ }\textbf {\bibinfo
  {volume} {68}},\ \bibinfo {pages} {879}} (\bibinfo {year} {1955})\BibitemShut
  {NoStop}%
\bibitem [{\citenamefont {Azbel}(1964)}]{Azbel:1964_SI}%
  \BibitemOpen
  \bibfield  {author} {\bibinfo {author} {\bibfnamefont {M.~Y.}\ \bibnamefont
  {Azbel}},\ }\href@noop {} {\bibfield  {journal} {\bibinfo  {journal} {Zh.
  Eksp. Teor. Fiz.}\ }\textbf {\bibinfo {volume} {46}},\ \bibinfo {pages}
  {929}} (\bibinfo {year} {1964}),\ \bibinfo {note} {[Sov. Phys. JETP
  \textbf{19}, 634 (1964)]}\BibitemShut {NoStop}%
\bibitem [{\citenamefont {Hofstadter}(1976)}]{Hofstadter:1976_SI}%
  \BibitemOpen
  \bibfield  {author} {\bibinfo {author} {\bibfnamefont {D.~R.}\ \bibnamefont
  {Hofstadter}},\ }\href {\doibase 10.1103/PhysRevB.14.2239} {\bibfield
  {journal} {\bibinfo  {journal} {Phys. Rev. B}\ }\textbf {\bibinfo {volume}
  {14}},\ \bibinfo {pages} {2239}} (\bibinfo {year} {1976})\BibitemShut
  {NoStop}%
\bibitem [{\citenamefont {Hatsugai\ and\ Kohmoto}(1990)\citenamefont
  {Hatsugai}\ and\ \citenamefont {Kohmoto}}]{Hatsugai:1990_SI}%
  \BibitemOpen
  \bibfield  {author} {\bibinfo {author} {\bibfnamefont {Y.}~\bibnamefont
  {Hatsugai}}\ \bibnamefont {and}\ \bibinfo {author} {\bibfnamefont
  {M.}~\bibnamefont {Kohmoto}},\ }\href {\doibase 10.1103/PhysRevB.42.8282}
  {\bibfield  {journal} {\bibinfo  {journal} {Phys. Rev. B}\ }\textbf {\bibinfo
  {volume} {42}},\ \bibinfo {pages} {8282}} (\bibinfo {year}
  {1990})\BibitemShut {NoStop}%
\bibitem [{\citenamefont {Lohse et~al.}(2016)\citenamefont {Lohse},
  \citenamefont {Schweizer}, \citenamefont {Zilberberg}, \citenamefont
  {Aidelsburger},\ and\ \citenamefont {Bloch}}]{Lohse:2016_SI}%
  \BibitemOpen
  \bibfield  {author} {\bibinfo {author} {\bibfnamefont {M.}~\bibnamefont
  {Lohse}}, \bibinfo {author} {\bibfnamefont {C.}~\bibnamefont {Schweizer}},
  \bibinfo {author} {\bibfnamefont {O.}~\bibnamefont {Zilberberg}}, \bibinfo
  {author} {\bibfnamefont {M.}~\bibnamefont {Aidelsburger}},\ \bibnamefont
  {and}\ \bibinfo {author} {\bibfnamefont {I.}~\bibnamefont {Bloch}},\ }\href
  {http://dx.doi.org/10.1038/nphys3584} {\bibfield  {journal} {\bibinfo
  {journal} {Nat. Phys.}\ }\textbf {\bibinfo {volume} {12}},\ \bibinfo {pages}
  {350}} (\bibinfo {year} {2016})\BibitemShut {NoStop}%
\bibitem [{\citenamefont {Roux et~al.}(2008)\citenamefont {Roux}, \citenamefont
  {Barthel}, \citenamefont {McCulloch}, \citenamefont {Kollath}, \citenamefont
  {Schollw\"ock},\ and\ \citenamefont {Giamarchi}}]{Roux:2008_SI}%
  \BibitemOpen
  \bibfield  {author} {\bibinfo {author} {\bibfnamefont {G.}~\bibnamefont
  {Roux}}, \bibinfo {author} {\bibfnamefont {T.}~\bibnamefont {Barthel}},
  \bibinfo {author} {\bibfnamefont {I.~P.}\ \bibnamefont {McCulloch}},
  \bibnamefont {et~al.},\ }\href {\doibase 10.1103/PhysRevA.78.023628}
  {\bibfield  {journal} {\bibinfo  {journal} {Phys. Rev. A}\ }\textbf {\bibinfo
  {volume} {78}},\ \bibinfo {pages} {023628}} (\bibinfo {year}
  {2008})\BibitemShut {NoStop}%
\bibitem [{\citenamefont {Kraus et~al.}(2013)\citenamefont {Kraus},
  \citenamefont {Ringel},\ and\ \citenamefont {Zilberberg}}]{Kraus:2013_SI}%
  \BibitemOpen
  \bibfield  {author} {\bibinfo {author} {\bibfnamefont {Y.~E.}\ \bibnamefont
  {Kraus}}, \bibinfo {author} {\bibfnamefont {Z.}~\bibnamefont {Ringel}},\
  \bibnamefont {and}\ \bibinfo {author} {\bibfnamefont {O.}~\bibnamefont
  {Zilberberg}},\ }\href {\doibase 10.1103/PhysRevLett.111.226401} {\bibfield
  {journal} {\bibinfo  {journal} {Phys. Rev. Lett.}\ }\textbf {\bibinfo
  {volume} {111}},\ \bibinfo {pages} {226401}} (\bibinfo {year}
  {2013})\BibitemShut {NoStop}%
\bibitem [{\citenamefont {Widera et~al.}(2005)\citenamefont {Widera},
  \citenamefont {Gerbier}, \citenamefont {F\"olling}, \citenamefont {Gericke},
  \citenamefont {Mandel},\ and\ \citenamefont {Bloch}}]{Widera:2005_SI}%
  \BibitemOpen
  \bibfield  {author} {\bibinfo {author} {\bibfnamefont {A.}~\bibnamefont
  {Widera}}, \bibinfo {author} {\bibfnamefont {F.}~\bibnamefont {Gerbier}},
  \bibinfo {author} {\bibfnamefont {S.}~\bibnamefont {F\"olling}}, \bibnamefont
  {et~al.},\ }\href {\doibase 10.1103/PhysRevLett.95.190405} {\bibfield
  {journal} {\bibinfo  {journal} {Phys. Rev. Lett.}\ }\textbf {\bibinfo
  {volume} {95}},\ \bibinfo {pages} {190405}} (\bibinfo {year}
  {2005})\BibitemShut {NoStop}%
\end{thebibliography}
